\pdfoutput=1  

\documentclass[aps,prd,superscriptaddress,floatfix,nofootinbib,preprintnumbers,eqsecnum,twocolumn]{revtex4-1}

\usepackage[utf8]{inputenc}
\usepackage[dvips]{graphicx}
\usepackage[dvipsnames]{xcolor}

\usepackage{amsmath,float,graphicx,placeins,amssymb,multirow,pgfplots,pgfplotstable}
\usepackage{empheq}
\usepackage{tikz}
\usepackage{comment}
\usepackage{enumerate,slashed,cancel,comment,bbm,rotating,placeins,mathtools,multirow,hhline}
\usepackage[normalem]{ulem}
\usepackage{array}

\usepackage{epstopdf}
\usepackage{hyperref}
\usepackage{mathrsfs}
\usepackage{amssymb}
\usepackage{physics}
\usepackage{booktabs}

\usepackage{color}
\usepackage{relsize}
\usepackage{graphics}
\usepackage{epstopdf}
\usepackage{hyperref}
\usepackage{mathrsfs}
\usepackage{amssymb}
\usepackage{physics}
\usepackage{booktabs}

\usepackage[normalem]{ ulem }
\usepackage{amsthm}
\usepackage{amsmath}
\usepackage{cancel}
\usepackage{float}
\usepackage{fontawesome} 
\usepackage[caption=false]{subfig}
\usepackage{tabularx,ragged2e} 
 \newcolumntype{L}{>{\RaggedRight\arraybackslash}X}
 
\colorlet{mylinkcolor}{ForestGreen}
\colorlet{mycitecolor}{Red}
\colorlet{myurlcolor}{violet}

\usepackage{lipsum}
\usepackage{diagbox}
\usetikzlibrary{shapes.geometric,arrows}
\graphicspath{}

\def\beq{\begin{equation}}
\def\eeq{\end{equation}}
\def\beqn{\begin{eqnarray}}
\def\eeqn{\end{eqnarray}}

\hypersetup{
 linkcolor = {blue!80!black},
 citecolor = {blue!70!black},
 urlcolor = {blue!80!black},
 colorlinks = true
}
\def\barOmega{{\overline{\Omega}}}

\def\ie{{\it i.e.}\/}
\def\eg{{\it e.g.}\/}

\def\IR{\relax{\rm I\kern-.18em R}}
 \font\cmss=cmss10 \font\cmsss=cmss10 at 7pt
\def\IQ{\relax{\rm I\kern-.18em Q}}
\def\IZ{\relax\ifmmode\mathchoice
 {\hbox{\cmss Z\kern-.4em Z}}{\hbox{\cmss Z\kern-.4em Z}}
 {\lower.9pt\hbox{\cmsss Z\kern-.4em Z}}
 {\lower1.2pt\hbox{\cmsss Z\kern-.4em Z}}\else{\cmss Z\kern-.4em Z}\fi}

\def\M0{M^{(0)}}
\def\f0{f^{(0)}}
\def\t0{t^{(0)}}
\def\X0{X^{(0)}}
\def\a0{a^{(0)}}
\def\H0{H^{(0)}}

\def\nn{\nonumber}

\def\MD{M_D}

\pgfplotsset{compat=1.17}
\begin{document}


\title{ Primordial Black Holes and their Mass Spectra:\\
The Effects of Mergers and Accretion within Stasis Cosmologies}

\def\andname{\hspace*{-0.5em}} 

\author{Keith R. Dienes}
 \email[Email address: ]{dienes@arizona.edu}
 \affiliation{Department of Physics, University of Arizona, Tucson, AZ 85721 USA}
 \affiliation{Department of Physics, University of Maryland, College Park, MD 20742 USA}
\author{Lucien Heurtier}
\email[Email address: ]{lucien.heurtier@kcl.ac.uk}
\affiliation{Theoretical Particle Physics and Cosmology, King’s College London,
Strand, London WC2R 2LS, United Kingdom}
\author{Fei Huang}
\email[Email address: ]{fei.huang@weizmann.ac.il}
 \affiliation{Department of Particle Physics and Astrophysics, Weizmann Institute of Science, Rehovot 7610001, Israel}
  \author{Tim M.P. Tait}
 \email[Email address: ]{ttait@uci.edu}
 \affiliation{Department of Physics and Astronomy, University of California, Irvine, CA  92697  USA}
 \author{Brooks Thomas}
 \email[Email address: ]{thomasbd@lafayette.edu}
 \affiliation{Department of Physics, Lafayette College, Easton, PA  18042 USA}

\begin{abstract}
A variety of processes in the very early universe can give rise to a population of 
primordial black holes (PBHs) with an extended mass spectrum.  For certain mass spectra 
of this sort, it has been shown that the evaporation of these PBHs into 
radiation can drive the universe toward an epoch of cosmological stasis which 
can persist for a significant number of $e$-folds of cosmological expansion.
However, in general, the initial mass spectrum which characterizes a population of PBHs 
at the time of production can subsequently be distorted by processes such as mergers and 
accretion.  In this paper, we examine the effects that these processes have on 
the spectra that lead to a PBH-induced stasis.  Within such stasis models, we find that 
mergers have only a negligible effect on these spectra within the regime of interest for 
stasis.  We likewise find that the effect of accretion is negligible in many cases
of interest.  However, we find that the effect of accretion on the PBH mass spectrum is 
non-negligible in situations in which this spectrum is particularly broad.  In such 
situations, the stasis epoch is abridged or, in extreme cases, does not occur at all. 
Thus accretion plays a non-trivial role in constraining the emergence of stasis within 
scenarios which lead to extended PBH mass spectra.
\end{abstract}
\maketitle

\tableofcontents


\section{Introduction\label{sec:intro}}


Primordial black holes (PBHs) --- \ie, black holes which are not produced via 
stellar dynamics --- can be generated by a number of processes within the early 
universe (for reviews, see, \eg, Refs.~\cite{Green:2014faa,Escriva:2022duf}).  
The existence of such objects
can have a variety of implications for early-universe cosmology.  PBHs can contribute 
to the present-day dark-matter abundance~\cite{Hawking:1971ei,Chapline:1975ojl} and 
serve as engines for the production of dark-matter particles~\cite{Baumann:2007yr,
Fujita:2014hha,Lennon:2017tqq,Allahverdi:2017sks,Morrison:2018xla,Hooper:2019gtx,
Masina:2020xhk,Baldes:2020nuv} or of heavy, unstable particles whose baryon-number-violating 
decays contribute to the generation of a baryon asymmetry for the
universe~\cite{Baumann:2007yr,Fujita:2014hha,Morrison:2018xla}.  

PBHs can also potentially have an impact on the expansion history of the universe 
(for reviews, see, \eg, Ref.~\cite{Batell:2024dsi}).  For example, a population of PBHs 
can come to dominate the energy density of the universe prior to the Big-Bang 
nucleosynthesis (BBN) epoch, thereby giving rise to an early matter-dominated era.  
Alternatively, a cosmological population of black holes with an extended mass spectrum 
can even give rise to a situation in which the abundances of matter and radiation remain 
effectively constant for a significant duration~\cite{Barrow:1991dn,Dienes:2022zgd}.  
In other words, such a population of PBHs furnishes a realization of 
{\it cosmological stasis}\/~\cite{Dienes:2021woi,Dienes:2023ziv} --- a phenomenon wherein 
energy density is transferred from cosmological energy components with smaller 
equation-of-state parameters to those with larger equation-of-state parameters in such 
a way as to precisely compensate for the differing manner in which the energy densities 
associated with these individual components are affected by cosmic expansion, thereby 
resulting in constant abundances for these different energy  components despite the 
expansion of the universe.  In this realization of the stasis phenomenon, PBH evaporation 
via the emission of Hawking radiation is the mechanism which facilitates the transfer of 
energy density from matter to radiation.  Indeed, within this realization of stasis, it 
can be shown that the stasis state is a global dynamical 
attractor~\cite{Barrow:1991dn,Dienes:2022zgd}, just as it is within the many other 
stasis cosmologies that have been shown to exist~\cite{Dienes:2021woi,Dienes:2023ziv,
Dienes:2024wnu,Halverson:2024oir,Barber:2024vui,Barber:2024izt,Huang:2025odd,Dienes:2025tox,
Long:2025wjw}.  Thus, even though their initial conditions may vary widely, such PBH cosmologies 
are inevitably drawn towards a stasis configuration in which the effects of Hawking radiation 
and cosmological expansion are precisely balanced against each other.  Moreover, realizations 
of stasis involving PBHs can give rise to distinctive phenomenological signatures, including 
characteristic imprints on the stochastic gravitational-wave
background~\cite{Dienes:2022zgd}.  

This stasis-inducing interplay between cosmological expansion and PBH evaporation  
is ultimately sensitive to the manner in which the differential number density 
$f_{\rm BH}(M,t)$ for the population of PBHs scales with the PBH mass $M$ across the 
extended PBH spectrum.  In situations in which these two processes are the only ones that 
have a non-negligible impact on the evolution of $f_{\rm BH}(M,t)$, the PBH mass spectrum 
evolves with time in a particularly simple manner.  Cosmological expansion reduces the 
physical number density of PBHs, thereby decreasing the overall {\it normalization}\/ of 
$f_{\rm BH}(M,t)$ but leaving the {\it shape}\/ of $f_{\rm BH}(M,t)$ intact.  By contrast,
over time, evaporation effectively converts the (matter) energy density associated with 
PBHs into radiation.  This process is actually quite abrupt --- far more so than, \eg, the 
exponential decay of a population of unstable particles --- in the sense that the vast 
majority of the initial mass of a given PBH is converted into radiation within a 
comparatively short time interval immediately prior to the time at which the PBH 
evaporates completely.  Evaporation therefore does not have a significant impact on the 
shape of $f_{\rm BH}(M,t)$ at values of $M$ significantly larger than the mass of the 
lightest PBH which has yet to evaporate at time $t$.  

By contrast, in situations in which additional processes have a non-negligible impact on 
the time-evolution of $f_{\rm BH}(M,t)$, the initial shape of this differential number density 
can be distorted over time.  Such distortions can potentially disrupt the dynamics which  
gives rise to stasis.  As it turns out, the two processes which have the greatest potential 
to alter the shape of $f_{\rm BH}(M,t)$ in this way are the accretion of material onto the 
PBHs and mergers between PBHs.

In this paper, we shall therefore examine the effects that mergers and accretion have on 
$f_{\rm BH}(M,t)$ in PBH-induced-stasis scenarios.  To do this, we shall consider the 
PBH-induced stasis model presented in Ref.~\cite{Dienes:2022zgd} and identify those 
parameter-space regions associated with this model within which mergers and accretion 
are unimportant across the entire PBH mass spectrum.  These are therefore regions of 
parameter space within which the stasis-inducing dynamics is essentially undisturbed.  
We shall also identify those regions in which these effects might lead to more significant 
distortions.

Our primary results can be summarized as follows.  As we shall find, the effects of 
mergers on $f_{\rm BH}(M,t)$ are negligible across the entire region of parameter 
space relevant for stasis.  Thus the effects of mergers are ultimately not a concern 
for the stasis-inducing dynamics in Ref.~\cite{Dienes:2022zgd}.  By contrast, whether 
or not the effects of accretion have a significant impact on $f_{\rm BH}(M,t)$ depends 
on the location in parameter space.  Throughout the vast majority of the parameter space 
relevant for stasis --- and indeed the majority of that parameter space wherein the prospects 
for detecting the distinctive phenomenological signatures identified in 
Ref.~\cite{Dienes:2022zgd} are particularly auspicious --- the effects of accretion are 
also negligible.  Thus, throughout much of the stasis-producing parameter space associated 
with the model of Ref.~\cite{Dienes:2022zgd}, the dynamics leading to stasis is largely 
unaffected by either mergers or accretion.  However, within certain regimes --- in particular, 
those in which the PBH mass spectrum is almost maximally broad or in which the PBHs collectively 
dominate the energy density of the universe for a significant duration prior to stasis --- 
accretion can have a more significant impact on $f_{\rm BH}(M,t)$. In such situations, the 
duration of the stasis epoch can be reduced.  Indeed, in extreme cases of this sort, we find 
that stasis might not occur at all.

This paper is organized as follows.  In Sect.~\ref{sec:pbhstasis}, we briefly review the 
model of PBH-induced stasis presented in Ref.~\cite{Dienes:2022zgd} and the cosmological 
expansion history associated with this model.  In Sect.~\ref{sec:binarycapture}, we then 
examine the process through which pairs of PBHs form binaries within the context of this 
cosmology.  In Sect.~\ref{sec:coalescence}, we consider the statistical properties of such PBH 
binaries and derive an expression for the overall rate at which the PBHs within these binaries  
coalesce into heavier black holes.  In Sect.~\ref{sec:results}, we then investigate the impact 
that PBH mergers have on $f_{\rm BH}(M,t)$ within the parameter-space regions 
relevant for stasis.  In Sect.~\ref{sec:accretion}, we consider the effects of accretion on 
the evolution of $f_{\rm BH}(M,t)$ and assess how these effects impact the duration of 
stasis within different stasis-producing parameter-space regions.  Finally, in 
Sect.~\ref{sec:conclusions}, we conclude with a summary of our main results and a discussion 
of possible directions for future work.  Appendix~\ref{app:BinConfig} provides a derivation of  
the probability densities associated with different configurations of PBHs which give rise 
to binaries with different total masses and geometries.


\section{Preliminaries\label{sec:pbhstasis}}


A population of non-relativistic Schwarzschild PBHs in the early universe may be 
characterized at a given cosmological time $t$ by their differential number density 
$f_{\rm BH}(M,t)$ per unit PBH mass $M$.  In general, the Boltzmann equation which 
describes the manner in which $f_{\rm BH}(M,t)$ evolves may be written as
\begin{eqnarray}
  \frac{df_{\rm BH}(M,t)}{dt} &\,=\,& -3Hf_{\rm BH}(M,t) \nonumber \\
    & & +\, \frac{\partial f_{\rm BH}(M,t)}{\partial M} 
    \left[\left(\frac{dM}{dt}\right)_e + 
    \left(\frac{dM}{dt}\right)_{\rm acc} \right]
    \nonumber \\
    & & +\, \big[\Gamma_+(M,t) - \Gamma_-(M,t)\big] f_{\rm BH}(M,t) ~,
    \nonumber \\
  \label{eq:MNote_urBoltz}
\end{eqnarray}
where $H \equiv \dot{a}/a$ is the Hubble parameter, where $(dM/dt)_e$ and 
$(dM/dt)_{\rm acc}$ respectively denote the rates at which $M$ itself changes due to 
the emission of Hawking radiation and due to the accretion of material around it, and 
where the products $\Gamma_+(M,t) f_{\rm BH}(M,t)$ and $\Gamma_-(M,t) f_{\rm BH}(M,t)$ 
respectively denote the rates at which $f_{\rm BH}(M,t)$ changes due to the production 
of PBHs of mass $M$ via the mergers of lighter PBHs and due to the disappearance of 
PBHs of mass $M$ via their mergers with other PBHs.  Parametrizing the 
merger rates in terms of $\Gamma_\pm(M,t)$, which have units of inverse time, facilitates 
comparison between these quantities and $H$, thereby providing a transparent method of 
assessing whether mergers have a significant effect on the differential number density
of PBHs at any given point in the cosmological history.  In particular, at any time at which 
either $\Gamma_+(M,t)$ or $\Gamma_-(M,t)$ is comparable to or larger than $H$ for any particular
value of $M$, the effect of mergers on $f(M,t)$ at that value of $M$ cannot be neglected.  
By contrast, at any time at which $\Gamma_+(M,t)$ and $\Gamma_-(M,t)$ are both significantly 
smaller than $H$ for a particular value of $M$, the effect of mergers on $f_{\rm BH}(M,t)$ 
at that value of $M$ is insignificant.

The quantities $\Gamma_+(M,t)$ and $\Gamma_-(M,t)$, to which we shall henceforth refer as the 
PBH production and destruction rates from mergers, respectively, both depend non-trivially 
on the full functional form of $f_{\rm BH}(M,t)$.  Indeed, mergers between PBHs of mass $M$ and 
PBHs of any mass contribute to $\Gamma_-(M,t)$, while mergers between any two PBHs whose masses 
sum to $M$ contribute to $\Gamma_+(M,t)$.  The functional form of $f_{\rm BH}(M,t)$ at any given 
time in turn depends both on its initial form $f_{\rm BH}(M,t_i)$ at the time $t_i$ at which the 
cosmological population of PBHs in question is effectively established and on the expansion 
history of the universe at times between $t_i$ and $t$.  Thus, before we begin to evaluate 
$\Gamma_\pm(M,t)$ within the context of PBH-induced stasis scenario put forth in 
Ref.~\cite{Dienes:2022zgd}, we briefly review the non-standard expansion history 
associated with this scenario and the properties of the corresponding cosmological 
population of PBHs.

\subsection{Mass Spectrum\label{sec:spectrum}}

One mechanism through which a cosmological population of PBHs with an extended mass 
spectrum can be generated within the early universe is via the collapse of primordial 
density perturbations after inflation.  The relationship between the initial mass $M_i$ of 
a PBH and the time $t_p(M_i)$ at which it is produced in this way is given by
(for reviews, see, \eg, Ref.~\cite{Byrnes:2021jka})
\begin{equation}
  M_i ~=~ \frac{4\pi}{3}\gamma \frac{\rho_{\rm crit}(t_p)}{H^3(t_p)}~, 
  \label{eq:MNote_PBHFormTime}
\end{equation}
where $H(t)$ is the Hubble parameter at time $t$ and where $\gamma$ is an $\mathcal{O}(1)$ 
proportionality factor which depends on the details of the collapse dynamics.
In what follows, we take $\gamma = 1$, though
we emphasize that our results are not particularly sensitive to the value of $\gamma$.
Since Eq.~(\ref{eq:MNote_PBHFormTime}) implies that $t_p(M_i)$ is a monotonically increasing 
function of $M_i$, it follows that the production of PBHs ends at the time 
$t_i \equiv t_p(M_{\rm max})$ at which the PBHs with the maximum initial mass $M_{\rm max}$ 
are produced.  

Once a PBH is produced, it begins losing mass via the emission of Hawking radiation.
The rate of change of $M$ due to evaporation is~\cite{MacGibbon:1990zk,MacGibbon:1991tj}
\begin{equation}
  \left(\frac{d M}{d t}\right)_e~\equiv~ -\varepsilon(M)\frac{M_P^4}{M^2}~,
  \label{eq:dMBHdt}
\end{equation}
where $M_P \equiv G^{-1/2}$ is the Planck mass, defined in terms of Newton's constant 
$G$, and where the evaporation function 
$\varepsilon(M)$ characterizes how this rate varies with $M$.  As discussed in 
Ref.~\cite{Dienes:2022zgd}, $\varepsilon(M) \approx \varepsilon$ is effectively independent 
of $M$ within our regime of interest for stasis.  For such a mass-independent evaporation 
function, Eq.~(\ref{eq:dMBHdt}) implies that the time $\tau_e(M_i)$ at which a PBH of 
initial mass $M_i$ evaporates completely is
\begin{equation}
  \tau_e(M_i) ~=~ \frac{M_i^3}{3\varepsilon M_P^4}~.
  \label{eq:MNote_timePBHevap}
\end{equation}
Moreover, Eq.~(\ref{eq:dMBHdt}) also implies that the majority of this initial mass is lost 
at times comparable to $\tau_e(M_i)$.  It is therefore reasonable for us to adopt an 
``instantaneous-evaporation'' approximation wherein we take the mass $M(t)$ of a PBH at 
to be approximately equal to its initial mass $M_i$ at any time 
$t_i < t < \tau_e(M_i)$.  

We shall focus in what follows on the case in which the entire spectrum of PBHs is 
generated during a single cosmological epoch wherein $w = w_c$, where $w_c$ is a 
constant.  The initial number density $f_{\rm BH}(M_i,t_i)$ of PBHs per unit $M_i$ 
which results from gravitational collapse in this case takes the 
form~\cite{Carr:1975qj,Green:1997sz,Kim:1999iv,Bringmann:2001yp,Carr:2017jsz}
\begin{equation}
  f_{\rm BH}(M_i,t_i) ~=~ 
  \begin{cases}
    C M_i^{\alpha-1} & {\rm for}~ M_{\rm min}\leq M_i\leq M_{\rm max}\\
    0 & \text{otherwise}~,
  \end{cases}
  \label{eq:MNote_dist}
\end{equation}
where $\alpha$ is a power-law exponent, where $C$ is a normalization coefficient, and 
where $M_{\rm min}$ denotes the minimum mass
of PBHs within this population.  The value of the scaling exponent $\alpha$ 
which appears in Eq.~(\ref{eq:MNote_dist}) is related to $w_c$ by~\cite{Carr:1975qj}
\begin{equation}
  w_c ~\equiv~ -\frac{\alpha + 1}{\alpha + 3}~.
  \label{eq:MNote_w_c}
\end{equation} 

The range for $\alpha$ which is physically motivated and also gives rise to stasis 
is $-2 \leq \alpha < -1$, which corresponds to the range 
$0 < w_c \leq 1$~\cite{Dienes:2022zgd}.  For $\alpha$ within this range, the 
normalization coefficient $C$ in Eq.~(\ref{eq:MNote_dist}) is
\begin{equation}
  C ~=~ \frac{(\alpha+1)\rho_{{\rm BH},i}}
    {M_{\rm max}^{\alpha +1} - M_{\rm min}^{\alpha + 1}}~,   
\end{equation}
where $\rho_{{\rm BH},i}$ is the total initial abundance of PBHs at $t = t_i$.  
The values of $M_{\rm min}$ and $M_{\rm max}$ are likewise constrained by 
phenomenological considerations.  Constraints on PBH decay from 
BBN~\cite{Carr:2009jm,Carr:2020gox,Keith:2020jww}  place an upper bound on $M_{\rm max}$
in situations in which $\rho_{{\rm BH},i}$ is sufficiently large that the PBHs come 
to dominate the energy density of the universe at any time,
while the upper bound from CMB data on the Hubble parameter during 
inflation~\cite{Planck:2018jri} in turn imposes a lower bound on $M_{\rm min}$.   
Given these constraints, we focus on combinations of $M_{\rm min}$ and $M_{\rm max}$ 
which satisfy
\begin{equation}
   0.1~{\rm g} ~\leq~ M_{\rm min} ~<~ M_{\rm max} ~\leq~ 10^9~{\rm g}~.
   \label{eq:MRange}
\end{equation}
Moreover, in what follows, we shall primarily be interested in
cases in which we also have $M_{\rm min} \ll M_{\rm max}$.

\subsection{Expansion History\label{sec:expansionhistory}}

In PBH-induced stasis scenarios of the sort discussed in Ref.~\cite{Dienes:2022zgd}, 
the portion of the expansion history between the end of inflation and the time 
$t_{\rm MRE}$ of matter-radiation equality can, to a very good approximation, be 
divided into four distinct epochs, during each of which the effective equation-of-state 
parameter $w$ is approximately constant.  The first of these is the epoch during which 
the PBHs are produced via gravitational collapse, wherein $w = w_c$.  Since 
$0 < w_c \leq 1$ for values of $\alpha$ relevant for stasis, as discussed below
Eq.~(\ref{eq:MNote_w_c}), $\rho_{\rm BH}$ grows during this epoch until the PBHs 
eventually come to dominate the energy density of the universe.  At this point, a 
second epoch begins --- one in which $w = 0$.  The time $t_f$ at which 
this the transition to this early matter-dominated epoch occurs for any given value 
of $w_c$ is determined by the initial energy density $\rho_{{\rm BH},i}$ at $t=t_i$.  

The third epoch is the stasis epoch itself, which effectively begins at the 
time $t_{\rm PBH}$ at which the lightest PBHs in the spectrum evaporate completely.  
It follows from Eq.~(\ref{eq:MNote_timePBHevap}) that
\begin{equation}
  t_{\rm PBH} ~=~ \frac{M_{\rm min}^3}{3\epsilon M_P^4}~.
  \label{eq:MNote_tPBH}
\end{equation}
During this stasis epoch, the effective equation-of-state parameter for the universe
is given by $w = \overline{w}$, where~\cite{Dienes:2022zgd}
\begin{equation}
    \overline{w} ~=~ -\frac{\alpha +1}{\alpha + 7}~.
\end{equation}
The stasis epoch lasts until the time $t_s$ at which the heaviest PBHs in the spectrum 
evaporate completely.  Thus, it follows from Eq.~(\ref{eq:MNote_timePBHevap}) that 
\begin{equation}
  t_s ~=~ \frac{M_{\rm max}^3}{3\epsilon M_P^4}~.
  \label{eq:MNote_ts}
\end{equation}
The fourth and final epoch prior to matter-radiation equality is the usual 
radiation-dominated epoch, which effectively begins at $t_s$ and lasts until 
$t_{\rm MRE}$.

\begin{figure}[b!]
  \centering
  \includegraphics[width=0.48\textwidth]{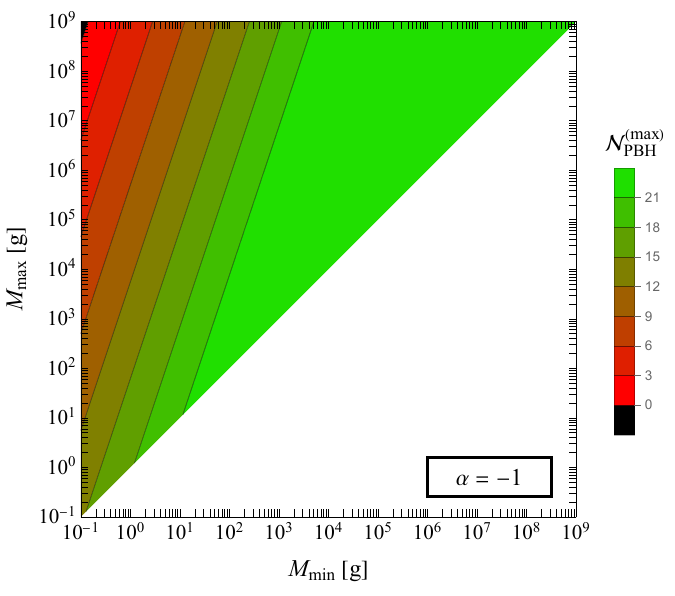}\\
  \caption{Contours of the maximum value $\mathcal{N}_{\rm PBH}^{(\rm max)}$ 
    of $\mathcal{N}_{\rm PBH}$ consistent with 
    our stasis cosmology, shown in the $(M_{\rm min},M_{\rm max})$-plane for $\alpha = -1$.  
    The black region of the plot is inconsistent with our cosmology for all 
    $\mathcal{N}_{\rm PBH} \geq 0$ and therefore excluded.  The corresponding upper bound 
    on $\mathcal{N}_{\rm PBH}$ is weaker for all $\alpha < -1$ throughout the entirety of
    the upper half-plane, though the value $\mathcal{N}_{\rm PBH}^{(\rm max)}$ is largely 
    insensitive to the choice of $\alpha$.
  \label{fig:NPBHMaxPlot}}
\end{figure}

The timescales $t_{\rm PBH}$ and $t_s$ associated with the beginning and end of
the stasis epoch --- as well as the equation-of-state parameters $w_c$ and $\overline{w}$
during the PBH-formation and stasis epochs, respectively --- are entirely determined 
by the parameters $M_{\rm min}$, $M_{\rm max}$, and $\alpha$ which characterize the
shape of the PBH mass spectrum.  By contrast, the timescale $t_f$ associated with
the beginning of the PBH-dominated epoch immediately prior to stasis also depends on
$\rho_{{\rm BH},i}$, which characterizes the overall normalization of this spectrum ---
or, equivalently, on the number of $e$-folds $\mathcal{N}_{\rm PBH}$ of cosmological 
expansion which take place during that PBH-dominated epoch.  The expansion history of 
our PBH-induced-stasis model is therefore completely characterized by the four free 
parameters $M_{\rm min}$, $M_{\rm max}$, $\alpha$, and $\mathcal{N}_{\rm PBH}$.

A mass spectrum of the simple form given in Eq.~(\ref{eq:MNote_dist}) arises
only in situations in which all PBHs form during a single cosmological epoch with a 
constant equation-of-state parameter.  Thus, our region of interest within the 
four-dimensional parameter space of this model is that wherein not only are the 
phenomenological constraints and consistency conditions on $M_{\rm min}$, $M_{\rm max}$, 
and $\alpha$ satisfied, but also wherein $\rho_{{\rm BH},i}$ is sufficiently small that 
the lighter PBHs within this spectrum do not collectively come to dominate the energy 
density of the universe before the heaviest PBHs have had a chance to form.  This upper 
bound on $\rho_{{\rm BH},i}$ can equivalently be expressed as an upper bound on 
$\mathcal{N}_{\rm PBH}$ for any combination of $M_{\rm min}$, $M_{\rm max}$, and $\alpha$.

In Fig.~\ref{fig:NPBHMaxPlot}, we show contours within the $(M_{\rm min},M_{\rm max})$-plane 
of the maximum value $\mathcal{N}_{\rm PBH}^{(\rm max)}$ of $\mathcal{N}_{\rm PBH}$ consistent 
with this upper bound.  The results shown in the figure correspond to the parameter 
choice $\alpha = -1$.  However, the value of $\mathcal{N}_{\rm PBH}^{(\rm max)}$ is not 
terribly sensitive to $\alpha$ for values of this parameter that lie within the range 
$-2 \leq \alpha < -1$ relevant for stasis.  Moreover, we note that this insensitivity 
is not merely a reflection of our taking $\gamma$ to be independent of $w_c$.  Indeed,
$\mathcal{N}_{\rm PBH}^{(\rm max)}$ is insensitive to $\alpha$ regardless of the manner in
which $\gamma$ depends non-trivially on $w_c$, provided of course that this presumably 
$\mathcal{O}(1)$ parameter in fact remains $\mathcal{O}(1)$ across this relevant range 
of $\alpha$.


\section{Binary Capture\label{sec:binarycapture}}


In general, PBH mergers proceed in two stages.  First, a pair of PBHs decouples from the
Hubble flow and forms a binary.  Second, the binary loses energy as a result of 
gravitational-wave emission and the two PBHs coalesce and merge into a single, more
massive black hole.  Thus, in order to determine the merger rate, we must consider how the 
characteristic timescales associated with both stages of this process depend on the 
properties of the merging PBHs and on their environment.  In doing so, we make extensive 
use of the formalism established in Ref.~\cite{Raidal:2017mfl} for evaluating the merger 
rate of a population of PBHs with an extended mass spectrum within the standard 
cosmology --- a formalism which we extend here to that of cosmologies with non-standard 
expansion histories.  Throughout the remainder of this section, we shall focus exclusively on 
binary formation.  We shall defer our discussion of the coalescence of PBHs within binaries 
after these binaries have formed until Sect.~\ref{sec:coalescence}.

In the Newtonian approximation, the physical proper distance $r$ between a pair of PBHs 
with masses $M_1$ and $M_2$ in a flat, Friedmann-Robertson-Walker (FRW) universe evolves 
according to the equation
\begin{equation}
  \ddot{r} ~=~ - G\frac{M_B}{r^2} + \frac{\ddot{a}}{a} r~,
  \label{eq:MNote_revolraw}
\end{equation}
where $M_B \equiv M_1 + M_2$ is the total mass of the system, where $a$ is the 
scale factor, and where a dot denotes a derivative with respect to the coordinate 
time $t$ in the cosmological background frame.  From the Friedmann acceleration 
equation, we have 
\begin{equation}
  \frac{\ddot{a}}{a} ~=~ -\frac{4\pi G}{3}(1 + 3w)\rho_{\rm crit} 
\medskip\smallskip
\end{equation}
where $\rho_{\rm crit}$ is the critical energy density of the universe and where $w$ is
the effective equation-of-state parameter for the universe as a whole.
Thus, Eq.~(\ref{eq:MNote_revolraw}) can be written as
\begin{equation}
  \ddot{r} ~=~ - G\frac{M_B}{r^2} - \frac{4\pi G}{3} (1+3w) \rho_{\rm crit} r~.
  \label{eq:MNote_revol}
\end{equation}

The time $t_B$ at which a pair of PBHs effectively decouples from the Hubble flow can 
be approximated as the time at which the gravitational term in Eq.~(\ref{eq:MNote_revol})
becomes larger in magnitude than the cosmological-expansion term.  In other words, $t_B$ 
is given implicitly by the solution to the equation
\begin{equation}
  M_B ~=~ \frac{4\pi}{3}(1+3w)\rho_{\rm crit}(t_B)r^3(t_B)~.
  \label{eq:MNote_deoupHflowraw}
\end{equation}
At times $t \lesssim t_B$, one may approximate $r(t)$ as evolving under the 
influence of cosmic expansion alone.  This implies that $r(t)$ evolves with the 
scale factor at such times according to the relation
\begin{equation}
  r(t) ~\approx~ \frac{a(t)}{a_\ast} r(t_\ast)~.
  \label{eq:MNote_roft}
\end{equation} 
In other words, the {\it comoving}\/ distance $x \equiv r(t)/a(t)$ between a pair of 
PBHs remains effectively constant until $t \sim t_B$.

The manner in which the critical density changes with $a(t)$ at any given time
depends on the abundances and equation-of-state parameters of the individual 
cosmological energy components present at that time.     
In what follows, we shall approximate the value of $w$ during each of these 
epochs as effectively constant and approximate the brief transition periods between 
them as instantaneous.  During any individual epoch within which $w$ remains constant,
the critical density evolves with $a$ according to the relation
\begin{equation}
  \rho_{\rm crit}(t) ~=~ \rho_{\rm crit}(t_\ast)
    \left[\frac{a(t)}{a_\ast}\right]^{-3(1+w)}~,
  \label{eq:MNote_rhocritevol}
\end{equation}
where $t_\ast$ is some fiducial time during that epoch and where 
$a_\ast \equiv a(t_\ast)$.
Combining this expression with the expression for $r(t)$ in Eq.~(\ref{eq:MNote_roft}), 
we find that the critical density is well approximated by
\begin{widetext}
\begin{equation}
    \rho_{\rm crit}(t)r^3(t) ~\approx~ \frac{1}{2}\rho_{\rm MRE}
      r^3(t_{\rm MRE}) \times
      \begin{cases}
        1 & t_{\rm MRE} < t \\
        \left(\frac{a}{a_{\rm MRE}}\right)^{-1} 
          & t_s < t < t_{\rm MRE} \\
        \left(\frac{a_s}{a_{\rm MRE}}\right)^{-1}
          \left(\frac{a}{a_s}\right)^{-3\overline{w}} 
          &  t_{\rm PBH} < t < t_s \\
        \left(\frac{a_s}{a_{\rm MRE}}\right)^{-1}
          \left(\frac{a_{\rm PBH}}{a_s}\right)^{-3\overline{w}}
           &  t_f < t < t_{\rm PBH} \\
        \left(\frac{a_s}{a_{\rm MRE}}\right)^{-1}
          \left(\frac{a_{\rm PBH}}{a_s}\right)^{-3\overline{w}} 
          \left(\frac{a}{a_f}\right)^{-3w_c} 
           &  t_{\rm end} < t < t_f~,
      \end{cases}
  \label{eq:MNote_rhor3raw}
\end{equation}
\end{widetext}
where $\rho_{\rm MRE} \equiv \rho_{\rm crit}(t_{\rm MRE})$, where $t_{\rm end}$ is the time 
at which inflation ends and the PBH-formation epoch begins, and where the overall factor 
of $1/2$ accounts for the fact that radiation represents half of the critical density 
at matter-radiation equality.  We note that the quantity $\rho_{\rm crit}(t)r^3(t)$ has 
no dependence on $a$ and remains constant throughout the PBH-domination epoch because 
$\rho_{\rm crit} \propto a^{-3}$ during an epoch of matter domination.  This implies 
that a pair of PBHs which have negligible velocities in the cosmological background 
frame cannot undergo binary capture during the PBH-dominated epoch.  Indeed, this is
the case in the later matter-dominated era which follows
$t_{\rm MRE}$ as well~\cite{Raidal:2017mfl}.

Substituting the result in Eq.~(\ref{eq:MNote_rhor3raw}) into Eq.~(\ref{eq:MNote_deoupHflowraw}), 
we find that the value of the scale factor $a_B \equiv a(t_B)$ at the time that a pair of 
PBHs forms a binary is
\begin{equation}
  a_B ~=~ \begin{cases}
    \displaystyle a_{\rm MRE}\left(\frac{x}{\widetilde{x}}\right)^3
    & \displaystyle  1 > \frac{x}{\widetilde{x}} > Q_s \\
    \displaystyle \frac{a_{\rm PBH}}{Q_{\rm PBH}^{1/\overline{w}}} 
    \left(\frac{x}{\widetilde{x}}\right)^{1/\overline{w}}
    & \displaystyle Q_s > \frac{x}{\widetilde{x}} > Q_{\rm PBH} \\
    a_{\rm PBH} & \displaystyle Q_{\rm PBH} > \frac{x}{\widetilde{x}} > Q_f\\
    \displaystyle \frac{a_f}{Q_f^{1/w_c}}\left(\frac{x}{\widetilde{x}}\right)^{1/w_c} 
    & Q_f > \displaystyle \frac{x}{\widetilde{x}}
   \end{cases}
  \label{eq:MNote_aBfull}
\end{equation}
where 
\begin{equation}
  \widetilde{x} ~\equiv~
    \left(\frac{3M_B}{4\pi\rho_{\rm MRE}a_{\rm MRE}^3}\right)^{1/3}
\end{equation}
is a fiducial distance that depends on the overall mass $M_B$ of the binary
and where we have defined
\begin{eqnarray}
  Q_{\rm PBH} &\,\equiv\,& \left[\frac{2a_s^{1-3\overline{w}}a_{\rm PBH}^{3\overline{w}}}
    {(1+3\overline{w})a_{\rm MRE}}\right]^{1/3} \nonumber \\
  Q_f &\,\equiv\,& \left[\frac{2a_s^{1-3\overline{w}}a_{\rm PBH}^{3\overline{w}}}
    {(1+3w_c)a_{\rm MRE}}\right]^{1/3} \,=\,
    \left(\frac{1+3\overline{w}}{1+3w_c}\right)^{1/3}Q_{\rm PBH} \nonumber \\
  Q_s &\,\equiv\,& \left(\frac{a_s}{a_{\rm MRE}}\right)^{1/3}~.
\end{eqnarray}
These dimensionless quantities represent the normalized initial comoving separations 
$x/\widetilde{x}$ between two PBHs which decouple from the Hubble flow and form
a binary at $t_{\rm PBH}$, $t_f$ and $t_s$, respectively.  We note that $Q_{\rm PBH}$, 
$Q_f$, and $Q_s$ depend on the properties of the initial PBH mass spectrum as a whole, 
but not on the masses of the PBHs that constitute a particular binary. 

Since $w$ is constant during each epoch of our PBH-induced stasis cosmology, 
the relationship between $a$ and $t$ during any particular epoch is given by 
$a = a_\ast (t/t_\ast)^{2/(3+3w)}$, where once again $t_\ast$ is some fiducial time 
during that epoch.  Thus, it follows from Eq.~(\ref{eq:MNote_aBfull}) that $t_B$ itself 
is given by
\begin{equation}
  t_B ~=~ \begin{cases}
    \displaystyle t_{\rm MRE}\left(\frac{x}{\widetilde{x}}\right)^6
      & \displaystyle  Q_s < \frac{x}{\widetilde{x}} < 1 \\
    \displaystyle t_{\rm PBH} \left(
      Q_{\rm PBH}^{-1}\frac{x}{\widetilde{x}}\right)
      ^{(3+3\overline{w})/(2\overline{w})}
      & \displaystyle Q_{\rm PBH} < \frac{x}{\widetilde{x}} < Q_s \\
    t_{\rm PBH} & \displaystyle Q_f < \frac{x}{\widetilde{x}} < Q_{\rm PBH}\\
      \displaystyle t_f
      \left(Q_f^{-1}\frac{x}{\widetilde{x}}\right)^{^{(3+3w_c)/(2w_c)}} 
    & \displaystyle \frac{x}{\widetilde{x}} < Q_f ~. 
  \label{eq:MNote_tBfull}
  \end{cases}    
\end{equation}

Our expression for $a_B$ in Eq.~(\ref{eq:MNote_aBfull}) for $x$ values within the range 
$Q_f < x/\widetilde{x} < Q_{\rm PBH}$  deserves further comment.  On the one hand, the 
fact that we are approximating the transitions between epochs as instantaneous and the 
effective equation-of-state parameter $w$ as constant during each epoch leads to a situation 
in which $Q_f \neq Q_{\rm PBH}$.  On the other hand, the approximations that
we have employed in modeling the process of binary formation imply that a pair of PBHs 
cannot decouple from the Hubble flow during a matter-dominated epoch.  It therefore follows
that there should exist a well-defined threshold value of $x/\widetilde{x}$ for any given 
combination of model parameters below which a PBH pair decouples during the PBH-formation 
epoch and above which it decouples during the stasis epoch.  Given these considerations,
we take $a_B = a_{\rm PBH}$ for all PBH pairs with $Q_f < x/\widetilde{x} < Q_{\rm PBH}$,
such that they decouple immediately at the moment at which the PBH-dominated epoch concludes.  
We emphasize that this range of $x/\widetilde{x}$ is always quite narrow within our 
parameter-space regime of interest.  Indeed, for $\alpha$ within the range 
$-2\leq \alpha \leq -1$ relevant for stasis, we have 
${(2/5)^{1/3} \approx 0.737 < Q_f/Q_{\rm PBH} < 1}$.  As such, our choice of convention 
for handling PBH pairs with $x/\widetilde{x}$ within this narrow range does not have 
a significant impact on our results.

The expression in Eq.~(\ref{eq:MNote_aBfull}) holds at times well after the last PBHs
--- \ie, those with the largest initial masses --- are produced.  At very early times, 
additional subtleties arise related to the manner in which the PBHs in our scenario 
are produced.  First of all, a pair of PBHs clearly cannot form a binary before the time 
$t_p(M_i)$ at which the heavier of the two PBHs, whose mass is here denoted $M_i$, is 
produced.  It follows from Eq.~(\ref{eq:MNote_PBHFormTime}) that the scale factor 
$a_p(M_i) \equiv a(t_p(M_i))$ at this time is 
\begin{equation}
  a_p(M_i) ~=~ a_f\left[
    \frac{16\pi M_i^2 \rho_{\rm MRE}}{3\gamma^2 M_P^6}
    \frac{a_{\rm MRE}^4 a_s^{3\overline{w}-1}}{a_f^3a_{\rm PBH}^{3\overline{w}}}
    \right]^{1/(3+3w_c)}~.  
\end{equation}
It follows from Eq.~(\ref{eq:MNote_aBfull}) that the requirement that $a_B > a_p(M_i)$ 
can be expressed as a constraint on the initial comoving distance $x$ between the PBHs, 
which takes the form
\begin{equation}
  \frac{x}{\widetilde{x}} ~>~ Q_p~,
  \label{eq:MNote_aBgtap}
\end{equation}
where we have defined
\begin{equation}
  Q_p
   ~\equiv~ \left[\frac{2a_s^{1-3\overline{w}}a_{\rm PBH}^{3\overline{w}}a_p^{3w_c}(M_i)}
    {(1+3w_c)a_{\rm MRE}a_f^{3w_c}}
    \right]^{1/3}~=~Q_f\left(\frac{a_p}{a_f}\right)^{w_c}~.
  \label{eq:MNote_aBgtap_defQp}
\end{equation}
In cases in which this constraint is not satisfied, the PBH pair would already nominally 
have decoupled from the Hubble flow prior to the time at which the heavier member is 
produced.

Moreover, since PBHs form in our scenario due to the collapse of density perturbations
whose comoving wavelength $\lambda_k = 2\pi/k \sim (aH)^{-1}$ is comparable to the
comoving Hubble radius, it is conceivable --- and in fact quite likely --- that 
the lighter member of any PBH pair which fails to satisfy the constraint in 
Eq.~(\ref{eq:MNote_aBgtap}) will already lie within the volume of space which collapses 
to form the heavier member of that pair at the moment it is initially
produced.  When this is the case, the mass of the lighter PBH is incorporated into the 
mass of the heavier one as it forms.  Indeed, the form of $f_{\rm BH}(M_i,t_i)$ in
Eq.~(\ref{eq:MNote_dist}) accounts for this effect~\cite{Carr:1975qj}.  Since the volume of
space which collapses to form a PBH of $M_i$ is a sphere whose radius is equal to
the Schwarzschild radius of that PBH, it follows that the lighter member of a would-be 
PBH pair can only be regarded as a distinct object after the heavier member is produced
when the condition $x > x_p(M_i,M_i') \equiv 2\max\{M_i,M_i'\} G/a_p$ is satisfied, 
where $M_i$ and $M_i'$ denote the masses of two PBHs.  Isolating the dependence of 
$x_p(M_i,M_i')$ on these masses and using the relation in Eq.~(\ref{eq:MNote_w_c}) to 
simplify the exponent, we find that
\begin{eqnarray}
    x_p(M_i,M_i') &\,=\,& \hat{x} 
      \left(\frac{\max\{M_i,M_i'\}}{M_P}\right)^{(1+3w_c)/(3+3w_c)} \nonumber \\
    &\,=\,& \hat{x} \left(\frac{\max\{M_i,M_i'\}}{M_P}\right)^{-\alpha/3}~, 
  \label{eq:xpdef}
\end{eqnarray}
where we have defined
\begin{equation}
  \hat{x} ~=~ \frac{2}{M_P}\left[
    \frac{3\gamma^2 M_P^4}{16\pi \rho_{\rm MRE}}
    \frac{a_f^{-3w_c}a_{\rm PBH}^{3\overline{w}}}{a_{\rm MRE}^4 a_s^{3\overline{w}-1}}
    \right]^{1/(3+3w_c)}~.    
\end{equation}

Equivalently, we note the condition $x > x_p(M_i,M_i')$ can be expressed as
\begin{equation}
  \frac{x}{\widetilde{x}} ~ > ~ \chi_{B}(M_i,M_i') Q_p~,
  \label{eq:MNote_SepAboveSchwarz}
\end{equation}
where
\begin{equation}
  \chi_{B}(M_i,M_i') ~ \equiv~ 
    \left[\frac{(1+3w_c)\gamma^2 \max\{M_i,M_i'\}}{M_B}\right]^{1/3}
  \label{eq:ChiDef}
\end{equation}
is a dimensionless $\mathcal{O}(1)$ quantity that depends both on the total mass of
the PBH pair and on the manner in which that mass is partitioned between the two PBHs.
We note that in deriving Eq.~(\ref{eq:ChiDef}), we have used the fact that $a_p$ can
be written in terms of $Q_p$ as
\beq
  a_p~=~\frac{1}{\widetilde{x}M_P^2Q_p}
    \left[\frac{8M_i^2 M_B}{\gamma^2(1+3w_c)}\right]^{1/3}~.
\eeq

Given the allowed range for $w_c$, we note that the allowed range for $\chi_B(M_i,M_i')$ is 
\begin{equation}
  \left(\frac{\gamma^2}{2}\right)^{1/3} ~<~ \chi_B(M_i,M_i') 
    ~<~ (4\gamma^2)^{1/3}~.
  \label{eq:ChiMMBoundforgamma} 
\end{equation} 
Thus, for $\gamma=1$, we have $2^{-1/3}<\chi_B<2^{2/3}$.
In cases in which $\chi_B(M_i,M_i') > 1$ for any particular combination of 
$M_i$, $M_i'$ and $M_B$, the lighter PBH in any PBH pair that would have decoupled 
from the Hubble flow prior to the formation of the heavier PBH 
(\ie, that has $a_B < a_p$ and thus $x/\widetilde{x} < Q_p$) always lies within the 
Schwarzschild radius of the heavier PBH and is therefore always incorporated into that 
heavier PBH at the moment it forms.  By contrast,  in cases in which $\chi_B(M_i,M_i') < 1$, 
there are values of $x$ for which the lighter PBH lies outside the Schwarzschild radius of 
the heavier PBH at this moment.  In such cases, the decoupling condition is satisfied 
immediately at the moment that $x = x_p(M_i,M_i')$ and the heavier PBH forms.

In order to express $a_B$ in terms of our fundamental model parameters $\alpha$, $M_{\rm max}$, 
$M_{\rm min}$, and $\mathcal{N}_{\rm PBH}$, we first note that, by definition,
\begin{equation}
  a_f ~=~ e^{-\mathcal{N}_{\rm PBH}} a_{\rm PBH}~.
\end{equation}
Moreover, during any cosmological epoch wherein $w$ is constant, 
$a$ and $t$ are related by
\begin{equation}
  \frac{a}{a_\ast} ~=~ \left(\frac{t}{t_\ast}\right)^{2/(3+3w)}~.
\end{equation} 
Thus, given Eqs.~(\ref{eq:MNote_tPBH}) and Eqs.~(\ref{eq:MNote_ts}) we have
\begin{equation}
  a_{\rm PBH} ~=~ 
    \left(\frac{M_{\rm min}}{M_{\rm max}}\right)^{2/(1+\overline{w})} a_s~.     
\end{equation}
Finally, since $w = 1/3$ between the end of stasis and the time of matter-radiation
equality, we have
\begin{equation}
  a_s ~=~ \left[\frac{\rho_{\rm crit}(t_{\rm MRE})}{\rho_{\rm crit}(t_s)}\right]^{1/4}
    a_{\rm MRE}~,
  \label{eq:MNote_asraw}
\end{equation}
where $a_{\rm MRE} = (1+z_{\rm MRE})^{-1}$, with $z_{\rm MRE} \approx 3402$~\cite{Planck:2018vyg}.  
Since the energy density at the 
end of the stasis epoch is
\begin{equation}
  \rho_{\rm crit}(t_s) ~=~ \frac{3H^2(t_s)}{8\pi G} 
    ~=~ \frac{3\epsilon^2 M_P^{10}}{2\pi (1+\overline{w})^2 M_{\rm max}^6}~,
\end{equation}
the expression in Eq.~(\ref{eq:MNote_asraw}) simplifies to
\begin{equation}
  a_s ~=~ \left[\frac{2\pi(1+\overline{w})^2\rho_{\rm MRE} M_{\rm max}^6}
    {3\epsilon^2 M_P^{10}}\right]^{1/4} a_{\rm MRE}~. 
\end{equation}

In a flat, $\Lambda$CDM universe, the critical density at matter-radiation equality is 
\begin{eqnarray}
    \rho_{\rm MRE} &=& 
      \frac{3H_{\rm now}^2 M_P^2}{8\pi}\bigg[\Omega_{\Lambda,{\rm now}} +
      \Omega_{M,{\rm now}}a_{\rm MRE}^{-3} \nonumber \\
        & &~+ \Omega_{\gamma,{\rm now}}a_{\rm MRE}^{-4} 
        \bigg]^{1/2}~,
\end{eqnarray}
where $H_{\rm now} \equiv H(t_{\rm now}) \approx 1.437\times 10^{-42}$~GeV and where 
$\Omega_{M,{\rm now}}\approx 0.315$, 
$\Omega_{\gamma,{\rm now}}\approx 5.38\times 10^{-5}$, and 
$\Omega_{\Lambda,{\rm now}}\approx 0.685$
denote the present-day abundances of matter, radiation, and vacuum energy,
respectively~\cite{Planck:2018vyg}.

\begin{figure}
  \centering
  \includegraphics[width=0.48\textwidth]{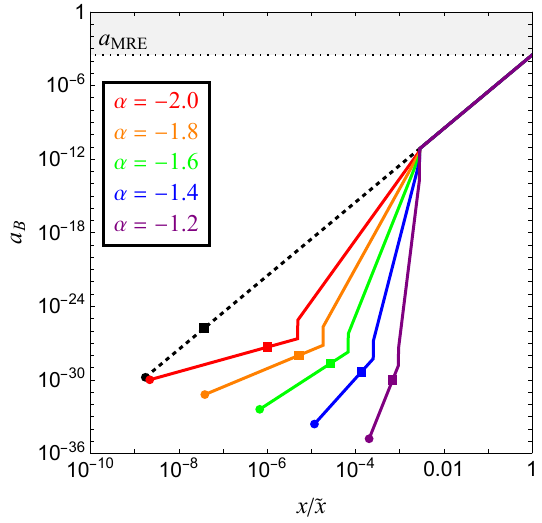} 
  \caption{The scale factor $a_B$ at the time at which a pair of PBHs decouple from the 
    Hubble flow and form a binary in a cosmology involving an epoch of PBH-induced stasis,
    shown for different values of $\alpha$ (solid colored curves).  All of the
    results shown correspond to the parameter choices $M_{\rm min} = 10$~g, 
    $M_{\rm max} = 10^9$~g, and $\mathcal{N}_{\rm PBH} = 5$.  The black dashed curve indicates 
    the corresponding result for $a_B$ in the standard cosmology.  The circular dot at the end 
    of each curve indicates the minimum value of $x/\widetilde{x}$ for a binary with 
    $M_B = 2M_{\rm min}$, given that a binary cannot form before its constituent PBHs are 
    produced via gravitational collapse.  The square that appears further to the right along 
    each curve indicates the corresponding minimum value of $x/\widetilde{x}$ for a binary 
    with $M_B = 2M_{\rm max}$. 
  \label{fig:MNote_aBPlotMax}}
\end{figure}

In Fig.~\ref{fig:MNote_aBPlotMax}, we plot $a_B$ as a function of 
$x/\widetilde{x}$ within our PBH-induced stasis cosmology for a variety of different
values of $\alpha$ (solid colored curves).  All of the results shown in the figure 
correspond to the parameter choices $M_{\rm min} = 10$~g, $M_{\rm max} = 10^9$~g, and 
$\mathcal{N}_{\rm PBH} = 5$.  The black dotted curve indicates the corresponding result 
for $a_B$ in the standard cosmology.  The circular dot at the end of each curve indicates 
the minimum value of $x/\widetilde{x}$ for a binary with $M_B = 2M_{\rm min}$, given that 
a binary cannot form before its constituent PBHs are produced.  The square that appears 
further to the right along each curve indicates the corresponding minimum value of 
$x/\widetilde{x}$ for a binary with $M_B = 2M_{\rm max}$.   

We observe from Fig.~\ref{fig:MNote_aBPlotMax} that the value of $a_B$ for a pair of PBHs with 
a given initial comoving separation $x$ in our PBH-induced stasis scenario is less than or 
equal to the corresponding result in the standard cosmology for all cases shown.  
This reflects the fact that $\rho_{\rm crit}$ decreases more rapidly with $a$ during a 
radiation-dominated epoch than it does during a stasis epoch with $0 < \overline{w} < 1/3$ 
or a PBH-dominated epoch with $w = 0$.  However, we note that while
it is typically the case that a pair of PBHs with a given value of $x$ decouple from the 
Hubble flow at an equal or smaller value of $a$ within our stasis cosmology than they would 
within the standard cosmology, there are exceptions to this general rule.  In particular, 
within the regime in which $\alpha < -3/2$, we have $w_c > 1/3$, which implies that $\rho_{\rm crit}$ 
decreases more slowly during a radiation-dominated epoch than it does during the PBH-formation epoch.  
As a result, within regions of model-parameter space in which $\alpha < -3/2$ and 
$\mathcal{N}_{\rm PBH}$ is small, very light black holes which are produced quite nearby 
each other can in fact form binaries slightly earlier within our stasis cosmology 
than they would within the standard cosmology.  Of course, it is also the case that PBHs 
which decouple after stasis ends, with comoving separations such that 
$x/\widetilde{x} > Q_s$, decouple at the same value of $a_B$ as they do in the 
standard cosmology.

\section{Coalescence\label{sec:coalescence}}

After they decouple from the Hubble flow, a pair of isolated PBHs which are 
initially at rest in the background frame will continue to travel along a 
straight-line path toward each other until they collide.  However, the 
presence of additional PBHs in the vicinity will disrupt the 
infall of those two PBHs, which will therefore typically form a binary
instead of colliding head-on.  Once such a binary forms, the two constituent 
PBHs lose energy via the emission of gravitational radiation and merge into a 
single black hole.  The properties of this binary --- and in particular its 
semi-major axis $r_a$ and semi-minor axis $r_b$ depend on the spatial distribution
of nearby PBHs.

For simplicity, following Refs.~\cite{Nakamura:1997sm,Ioka:1998nz}, we approximate the 
formation of these binaries as a three-body process wherein the orbital properties of the 
binary are determined by the gravitational influence of the single additional PBH with the
smallest comoving distance $y$ from the center of mass of the two PBHs which form the binary.  
Refinements which account for the effect of additional PBHs were considered 
in~\cite{Ali-Haimoud:2017rtz}.
In this approximation, the time that it takes for the merger of two 
PBHs with individual masses $M_1$ and $M_2$ whose infall is disrupted by a third 
PBH of mass $M_3$ to occur is~\cite{Peters:1964zz}
\begin{equation}
  \tau_m(M_1,M_2,M_3,x,y) ~\approx~ \frac{3}{85} 
    \frac{r_a^{-3}(x) r_b^7(x,y)}{G^3 M_1 M_2 (M_1+M_2)}~.
  \label{eq:MNote_tauMerge}
\end{equation}
The semi-major axis of the orbit turns out to be proportional to the initial physical 
distance $r_B = a_B x$ between the two PBHs in the binary at the time at which the system 
initially decouples from the Hubble flow, while the semi-minor axis is proportional 
to the product of the tidal force and the square of the free-fall time.  In particular,
one finds that
\begin{eqnarray}
  r_a(M_1,M_2, x) &\approx& \zeta_a a_B x \nonumber  \\
  r_b(M_1,M_2,M_3,x,y) &\approx& 
    \frac{2\zeta_bM_3r_a(x)}{(M_1+M_2)}\left(\frac{x}{y}\right)^3 ~,~~
\end{eqnarray}
where $\zeta_a$ and $\zeta_b$ are proportionality factors of order unity.  For the
case in which $M_1 = M_2$, a detailed analysis yields the values $\zeta_a \approx 0.4$ 
and $\zeta_b = 0.8$ for these proportionality factors~\cite{Ioka:1998nz,Ali-Haimoud:2017rtz}, 
though merger rates obtained in this case are not particularly sensitive to the precise 
values of $\zeta_a$ and $\zeta_b$~\cite{Sasaki:2016jop}.  Thus, while the values of 
$\zeta_a$ and $\zeta_b$ appropriate for a binary in which $M_1 \neq M_2$ may differ from
those appropriate for a binary in which $M_1 = M_2$, such differences will not 
have a drastic impact on our estimate for the merger rate.  Thus, following 
Ref.~\cite{Raidal:2017mfl}, we take $\zeta_a = \zeta_b = 1$.

On average, the differential number of PBHs at time $t$ per unit PBH mass per 
unit comoving radial distance which are located within a spherical shell of
comoving radius $x$ centered on a particular PBH of mass $M'$ can be expressed as 
\begin{equation}
  \frac{\partial^2 N(M,M',x)}{\partial M \partial x} ~=~ 
    4\pi x^2 a^3 f_{\rm BH}(M)\big[1+\xi(M,M',x)\big] ~,
  \label{eq:MNote_dNdMdx}
\end{equation}
where $\xi(M,M',x)$ is the two-point correlation function for the PBHs.  While both 
$\partial^2 N(M,M',x)/\partial M \partial x$ itself and $f_{\rm BH}(M)$ depend non-trivially 
on $t$, we have suppressed the time-dependence of both quantities in this equation
in order to reduce notational clutter.  

The functional form of the two-point correlation function $\xi(M,M',x)$ depends on the 
properties of the primordial density perturbations which collapse to produce the PBHs.  
This correlation function may, in general, depend on both $M$ and $M'$.
For simplicity, we shall henceforth assume that spatial fluctuation in the primordial energy
density are described by a Gaussian random field, in which case the spatial 
distribution of PBHs follows a Poisson distribution.  For a perfectly spatially 
homogeneous, Poisson-distributed population of PBHs, $\xi(M,M',x) = 0$ for all $x$, regardless
of the value of $M$.  
However, for more realistic distributions which depart from these idealized properties,
a number of additional effects give rise to a non-trivial dependence of $\xi(M,M',x)$ on $x$.
For the case of a monochromatic PBH mass spectrum, for example, it does not make sense 
for more than one PBH to form from gravitational collapse within the same horizon 
volume.  Thus, for comoving separations $x < x_p(M,M')$, it follows that 
$\xi(M,M',x) = 0$~\cite{Desjacques:2018wuu,Ballesteros:2018swv}.  For the case of an 
extended mass spectrum, the situation is more complicated, given that in this case lighter 
PBHs can form within regions of space which later collapse to form heavier PBHs.
Nevertheless, it has been shown that the same qualitative behavior is obtained in
this case~\cite{Auclair:2024jwj}.  Following Ref.~\cite{Raidal:2017mfl}, we 
approximate $\xi(M,M',x)$ as being independent of $x$ for $x > x_p(M,M')$ and take
\begin{equation}
  1 + \xi(M,M',x) ~\approx~ \delta_{\rm BH}\Theta\big(x-x_p(M,M')\big)~,  
\end{equation}
where $\delta_{\rm BH}$ is a constant and where $\Theta(x)$ denotes the Heaviside 
function of $x$.

It follows from Eq.~(\ref{eq:MNote_dNdMdx}) that the total number of PBHs of any mass located 
within a spherical comoving volume of radius $y$ centered on a PBH of mass $M'$ is
\begin{equation}
  N(M',y) ~\equiv~ \int_0^y dx \int_0^{\infty} dM\, 
    \frac{\partial^2 N(M,M',x)}{\partial M \partial x}~.
  \label{eq:MNote_Nofy}
\end{equation}
The upper limit of integration in the integral over $x$ is properly $y$ rather than 
$\max\{y,\widetilde{x}\}$ because $N(M',y)$ represents the number of PBHs 
within this spherical comoving volume, not the number of binaries.

The statistical properties of three-body configurations of PBHs may be expressed in terms of 
the quantities defined in Eqs.~(\ref{eq:MNote_dNdMdx}) and~(\ref{eq:MNote_Nofy}).
For example, we show in Appendix~\ref{app:BinConfig} that the joint probability density 
$\mathcal{P}_{{\rm n},{\rm nN}}(M_1,M_2,M_3,x,y)$ that the nearest neighbor of a particular PBH of 
mass $M_1$ will be located a comoving distance $x$ away from that first PBH, that the 
next-to-nearest neighbor of that same PBH will be located a distance $y$ away from that first PBH, 
and that these two PBHs will have masses $M_2$ and $M_3$, respectively, takes the form
\begin{eqnarray}
  && \!\!\!\!\! \mathcal{P}_{{\rm n},{\rm nN}}(M_1,M_2,M_3,x,y) ~=~ \nonumber \\
  && ~~~     e^{-N(M_1,y)} \frac{\partial^2 N(M_2,M_1,x)}{\partial M_2 \partial x}      
       \frac{\partial^2 N(M_3,M_1,y)}{\partial M_3 \partial y}~.~~~~~~~~
  \label{eq:MNote_PNnN}
\end{eqnarray}
We note that this joint probability density is independent of $M_1$.  The corresponding 
differential comoving number density $\widetilde{n}(M_1,M_2,M_3,x,y)$ of PBH nearest-neighbor 
pairs in which one PBH has mass $M_1$ and its nearest and next-to-nearest neighbor have these 
attributes is
\begin{eqnarray}
  && \widetilde{n}(M_1,M_2,M_3,x,y) ~=~ \frac{1}{2}a^3f_{\rm BH}(M_1) \nonumber \\
    && ~~~~~~~~~~~~~~~~~~~~~~~~\times \,
    \mathcal{P}_{{\rm n},{\rm nN}}(M_1,M_2,M_3,x,y)~,~~~~~~~~~
  \label{eq:MNote_ndiffofAllVars}
\end{eqnarray}
where the factor of $1/2$ accounts for the fact that there are two PBHs per binary and thus 
compensates for a double-counting.

Integrating the expression in Eq.~(\ref{eq:MNote_ndiffofAllVars}) over $x$ and $y$ while 
imposing the condition that the
total time $\tau_m(M_1,M_2,M_3,x,y) + t_B(M_1,M_2,x)$ that it takes for a pair 
of PBHs to form a binary and then coalesce
be equal to $t$, we obtain
\begin{eqnarray}
  &&\!\!\!\!\! \mathcal{R}_3(M_1,M_2,M_3) ~=~ \int_0^{\widetilde{x}} \!\! dx 
    \int_x^\infty\!\! dy 
    \Big[\widetilde{n}(M_1,M_2,M_3,x,y) \nonumber \\ 
    &&~~~~~~ \times\, \delta\big(t - \tau_m(M_1,M_2,M_3,x,y) 
     - t_B(M_1,M_2,x)\big) \Big]~, \nonumber \\ 
  \label{eq:MNote_MergerRateScriptR}
\end{eqnarray}
where the upper limit of integration in the integral over $x$ is properly $\widetilde{x}$
because PBH binaries do not form after matter-radiation equality.
We shall henceforth refer to $\mathcal{R}_3(M_1,M_2,M_3)$ as the differential three-body merger rate.  
Physically, it represents the differential contribution per comoving volume per unit $M_1$, 
$M_2$, and $M_3$ to the rate at which the differential comoving number density of PBH 
nearest-neighbor pairs with masses $M_1$ and $M_2$ whose next-to-nearest neighbor has mass 
$M_3$ decreases with $t$ due to binary mergers.  The quantity which is 
typically referred to in the literature as the differential PBH merger rate --- and to which 
we shall here refer as the differential two-body merger rate --- may be obtained simply by 
integrating this expression over all possible values for the mass of that next-to-nearest 
neighbor:
\begin{eqnarray}
  \mathcal{R}_2(M_1,M_2) &\equiv& \!\int_0^\infty\!\! dM_3\, \mathcal{R}_3(M_1,M_2,M_3) ~.
  \label{eq:MNote_MergerRateScriptR2}
\end{eqnarray}

The differential two-body merger rate gives rise to both the source and sink 
terms for $f_{\rm BH}(M)$ which appear in Eq.~(\ref{eq:MNote_urBoltz}).  Indeed, both 
$\Gamma_+(M,t)$ and $\Gamma_-(M,t)$ can be obtained from $\mathcal{R}(M_1,M_2)$ by direct 
integration.  In particular, these PBH production and destruction rates are given by
\begin{eqnarray}
  \Gamma_+(M) &=& \frac{1}{a^3f_{\rm BH}(M)} 
    \int_0^\infty dM_1 \int_0^\infty dM_2 \Big[ 
    \mathcal{R}_2(M_1,M_2) \nonumber \\ 
    & & ~~~\times\, \delta(M - M_1 - M_2) \Big] \nonumber \\
  \Gamma_-(M) &=& \frac{1}{a^3f_{\rm BH}(M)} 
    \int_0^\infty dM_2 \mathcal{R}_2(M,M_2)~,
  \label{eq:MNote_Gamma_pm}
\end{eqnarray}
where we have suppressed the explicit time-dependence in $\Gamma_\pm(x,t)$ in 
order to avoid notational clutter.   


\section{PBH Mergers: Results\label{sec:results}}


Our primary goal is to determine the region of parameter space within which  
$\Gamma_-(M,t)$ and $\Gamma_+(M,t)$ both remain sufficiently small 
until the end of the stasis epoch that the effect of mergers can be neglected.
Within this region, presuming for the moment that the effect of accretion can likewise 
be neglected, $f_{\rm BH}(M,t)$ evolves only as a consequence of evaporation and cosmological 
redshifting.  Indeed, if either of the conditions $\Gamma_+(M,t) \ll H$ and $\Gamma_+(M,t) \ll H$ 
is violated at any time $t$ within the range $t_i < t < \tau_e(M_{\rm max})$ for any $M$ within 
the range $M_{\rm min} \leq M \leq M_{\rm max}$, then the shape of $f_{\rm BH}(M,t)$ 
evolves non-trivially at all subsequent times and the dynamics which give rise to stasis 
are disturbed prior to the time at which the stasis epoch would otherwise have ended.

It is important to note, however, that even in situations in which these conditions are not 
satisfied for all $M$ within the range $M_{\rm min} \leq M \leq M_{\rm max}$, a stasis epoch 
is not necessarily precluded; rather, it may simply be abridged.
Mergers which involve a PBH of mass $M'$ can only affect the 
portion of $f_{\rm BH}(M,t)$ for which $M \geq M'$.  Moreover, since all of the PBHs 
are non-relativistic objects which behave like massive matter, mergers in and of 
themselves have no effect on the manner in which $\rho_{\rm BH}$ evolves with time. 
Indeed, the only way in which mergers affect the expansion history is by altering the
spectrum of evaporation times $\tau_e(M)$ for the population of PBHs.  However, since PBHs 
with larger masses evaporate later, mergers which involve a PBH of mass $M$ can only affect 
the expansion history at times $t \geq t_e(M)$.  Taken together, these observations imply
that if $\MD$ is the lowest value of $M$ for which the conditions
$\Gamma_+(M,t) \ll H$ and $\Gamma_-(M,t) \ll H$ are ever violated while 
$t_i < t < \tau_e(M_{\rm max})$, then distortions of $f_{\rm BH}(M,t)$ caused by mergers
only begin to affect the stasis dynamics at times $t > \tau_e(\MD)$. 
Thus, a population of PBHs with an initial mass spectrum of the form in Eq.~(\ref{eq:MNote_dist}) 
still gives rise to a stasis epoch, but one which ends at $t = \tau_e(\MD)$ rather than 
$t = \tau_e(M_{\rm max})$.  This stasis epoch is then followed by a transition period similar to that 
which arises in stasis scenarios involving towers of decaying particles~\cite{Dienes:2021woi}, 
wherein the effective equation-of-state parameter $w$ for the universe varies non-trivially with 
time.  This transition period ends once the heaviest PBHs ultimately generated by mergers evaporate 
completely, at which point the universe becomes radiation-dominated and remains so until 
$t_{\rm MRE}$.

For the purposes of identifying the region of our parameter space within which the effect of 
mergers on stasis are negligible and determining the value of $\MD$ 
for points which lie outside of this region, it is not necessary for us to determine 
precisely {\it how}\/ $f_{\rm BH}(M,t)$ behaves once either of the conditions
$\Gamma_+(M,t) \ll H$ or $\Gamma_-(M,t) \ll H$ is violated, but merely to determine 
{\it whether}\/ and {\it when}\/ that condition is violated.  Thus, proceeding in accord
with the assumption that the effect of accretion is also negligible --- an 
assumption which, as we shall see in Sect.~\ref{sec:accretion}, holds throughout the 
majority of our parameter-space region of interest  --- we may proceed by evaluating 
$\Gamma_+(M,t)$ and $\Gamma_-(M,t)$ as functions of $M$ and $t$ for the simple form of 
$f_{\rm BH}(M,t)$ which is obtained when only expansion and evaporation are important.

This form of $f_{\rm BH}(M,t)$ may be inferred from the observation that when only these 
two processes have a non-negligible impact on the differential number density, the comoving 
number density of PBHs with initial masses in the infinitesimal range from 
$M_i$ to $M_i + dM_i$ remains constant until these PBHs evaporate.  This implies that
\begin{equation}
  f_{\rm BH}(M,t)dM ~\approx ~ 
  \frac{a_i^3}{a^3} f_{\rm BH}(M_i,t_i)\Theta\Big(\tau_e(M_i) - (t - t_i) \Big)dM_i~, 
  \label{eq:MNote_fMt_fMit_equiv}
\end{equation}
where $a_i \equiv a(t_i)$.  Since ${M(t) \approx M_i\Theta(\tau_e(M_i)-(t-t_i))}$ 
in the instantaneous-evaporation approximation, it then follows that within this region 
of parameter space, $f_{\rm BH}(M,t)$ is approximately given by
\begin{eqnarray}
  f_{\rm BH}(M,t) &\,\approx\,& \frac{(\alpha + 1)\rho_{{\rm BH},i}M^{\alpha - 1}a_i^3}
    {(M_{\rm max}^{\alpha+1} -M_{\rm min}^{\alpha+1})a^3}  
    \Theta\big(M - M_{\rm cut}(t)\big) \nonumber \\ 
    && ~~\times\,\Theta(M_{\rm max} - M)~,~~~~~
  \label{eq:MNote_fMt_form}
\end{eqnarray}
where $M_{\rm cut}(t) \equiv \max\{M_{\rm min},\widetilde{M}(t)\}$ is the 
mass of the lightest PBH in the spectrum at time $t$ and where
\begin{equation}
  \widetilde{M}(t) ~\equiv~ \left[3\epsilon M_P^4(t-t_i)\right]^{1/3}~ \nn
\end{equation}
denotes the initial mass of the PBH whose evaporation time is $\tau_e(M_i) = t$.

For such a form of $f_{\rm BH}(M,t)$, the quantity defined in Eq.~(\ref{eq:MNote_dNdMdx}) 
reduces to 
\begin{eqnarray}
  \frac{\partial^2 N(M,M',x)}{\partial M \partial x} &\,=\,& 
      \frac{3\,\mathcal{C}_N x^2 M^{\alpha - 1}}{M_P^{\alpha-3}} 
        \Theta\big(x - x_p(M,M')\big)\nonumber \\ 
  && \,\times\,\Theta(M_{\rm max} - M) \Theta\big(M - M_{\rm cut}(t)\big)\,,
  \nonumber \\
  \label{eq:MNote_dNdMdxModel}
\end{eqnarray}
where we have defined the dimensionless coefficient
\begin{equation}
    \mathcal{C}_N ~\equiv~ 
      \frac{4\pi M_P^{\alpha-3} \rho_{{\rm BH},i} (\alpha +1)\delta_{\rm BH} a_i^3}
      {3(M_{\rm max}^{\alpha+1} - M_{\rm min}^{\alpha+1})}~. 
\end{equation}
Integrating this result over $x$ and $M$, we obtain
\begin{eqnarray}
  N(M',y) &=& \frac{\mathcal{C}_N}{M_P^{\alpha-3}}
    \int_{M_{\rm cut}(t)}^{M_{\rm max}} dM M^{\alpha - 1} 
    \left[y^3 - x_p^3(M,M')\right] \nonumber \\ 
    & &~~~~~~~~~~~~~~\times\,\Theta\big(y-x_p(M,M')\big)~.
\end{eqnarray}
Given the expression for $x_p(M,M')$ in Eq.~(\ref{eq:xpdef}), we observe that the 
Heaviside function appearing in this expression for $N(M',y)$ is equivalent to  
\begin{equation}
  \Theta\big(y-x_p(M,M')\big) ~=~ \Theta\big(M_*(y) -M\big)
    \Theta\big(M_*(y) - M'\big)~, \nonumber
\end{equation}
where have defined $M_*(y)\equiv M_P(y/\hat{x})^{-3/\alpha}$.
Thus, integrating over $M$, we find that until the time at which 
$\widetilde{M}(t) = M_{\rm max}$ and the entire population of PBHs 
has evaporated, our expression for $N(M',y)$ evaluates to
\begin{eqnarray}
  N(M',y) &=& 
    \frac{\mathcal{C}_N \hat{x}^3M_P^3}{\alpha} \Bigg[
    \frac{M_\uparrow^\alpha(y)-M_{\rm cut}^\alpha}{M_\ast^\alpha(y)}
    -\, \log\left(\frac{M_\uparrow^\alpha(y)}{M'^\alpha}\right) \nonumber \\
    & &~+\, \left(\frac{M_{\rm cut}}{M'}\right)^{\alpha} - 1 \Bigg]
    \Theta\big(M_\ast(y) - M'\big)~,
  \label{eq:MNote_NyModel}    
\end{eqnarray}
where we have defined $M_\uparrow \equiv \min\{M_{\rm max},M_\ast(y)\}$.

While the parametric dependence of $N(M',y)$ on $y$ is perhaps not
immediately transparent from Eq.~(\ref{eq:MNote_NyModel}), it is not
difficult to show that this quantity, which represents the number of 
PBHs within a spherical comoving volume of radius $y$, scales 
with $y$ in a manner which accords with this physical interpretation.
Indeed, we note that the contribution from the first term within the square brackets
on the right side of this equation typically dominates over the contribution 
from the remaining terms.  Thus, roughly speaking, $N(M',y)$ scales with $y$ as
\begin{equation}
  N(M',y) ~\sim~ 
    \begin{cases}
      \displaystyle\frac{M_{\rm cut}^\alpha}{M_\ast^\alpha(y)}
        \left(1- \frac{M_{\rm max}^\alpha}{M_{\rm cut}^\alpha}\right) 
        & M_\ast(y) > M_{\rm max} \\
      \displaystyle \frac{M_{\rm cut}^\alpha}{M_\ast^\alpha(y)} - 1 
        & M_\ast(y) < M_{\rm max}
    \end{cases}   
  \label{eq:NyInModelSimplified}
\end{equation}
at times $t \leq t_s$.  Regardless of the relationship between $M_\ast(y)$ and 
$M_{\rm max}$, this implies that $N(M',y)$ is a monotonically increasing function 
of $y$.  Moreover, Eq.~(\ref{eq:NyInModelSimplified}) also implies that within the 
regime in which $y$ is sufficiently large that $M_\ast(y) \gg M_{\rm cut}(t)$,
we may approximate $N(M',t)$ as
\begin{equation}
  N(M',t) ~\approx~ B_N(t) y^3~,
  \label{eq:NScalingWithyAndt}
\end{equation}
where the $y$-independent prefactor $B_N(t)$ in this expression scales differently with $t$ 
during different cosmological epochs.  Prior to the beginning of the stasis epoch,  
$M_{\rm cut} = M_{\rm min}$ and $B_N(t)$ is therefore independent of $t$.  By contrast, 
during the stasis epoch, $M_{\rm cut} = M_{\rm max}(t/t_s)^{1/3}$ and
$B_N(t)$ scales with $t$ as    
\begin{equation}
  B_N(t) ~\sim~ \begin{cases}
    t^{\alpha/3}\left[1- (t/t_s)^{-\alpha/3}\right] & M_\ast(y) > M_{\rm max} \\
    t^{\alpha/3}  & M_\ast(y) < M_{\rm max} ~.
  \end{cases}
\end{equation}
Regardless of the relationship between $M_\ast(y)$  and $M_{\rm max}$, this is well 
approximated by $B_N(t) \sim t^{\alpha/3}$ until the universe approaches the end of 
stasis.  Finally, for $t > t_s$, the entire population of PBHs has already evaporated, 
so $B_N(t) = 0$.  Physically, Eq.~(\ref{eq:NScalingWithyAndt}) indicates that within the 
regime in which $y$ is large, $N(M',y)$ is proportional to the volume of a comoving sphere 
of radius $y$, as expected.

Given the result in Eqs.~(\ref{eq:MNote_NyModel}), 
we are now equipped to evaluate $\mathcal{R}_3(M_1,M_2,M_3)$ in our PBH-induced 
stasis scenario.  Since $\tau_m(M_1,M_2,M_3,x,y)$ is a monotonically decreasing 
function of $y$, we find that the Dirac $\delta$-function appearing in 
Eq.~(\ref{eq:MNote_MergerRateScriptR}) may be expressed as
\begin{eqnarray}
  \delta\big(t - \tau_m(x,y)  - t_B(x) \big) 
  &\,=\,& \frac{y\delta\big(y-y_B(x,t)\big)}{21\tau_m(x,y)} \nonumber \\
  &\,=\,& \frac{y_B(x,t)}{21[t-t_B(x)]}\,\delta\big(y-y_B(x,t)\big)
  \nonumber \\
  &&~~~\times\, \Theta\big(t - t_B(x)\big) ~,  
  \label{eq:DeltFuncAlgebra}
\end{eqnarray}
where the Heaviside function is needed to ensure that only contributions 
from binaries which have formed by time $t$ contribute to the merger rate and
where we have defined
\begin{equation}
   y_B(x,t) ~\equiv~ x_B(t) 
   \left[\frac{x^{25}a^4_B(x)}{\widetilde{x}^{25} a^4_{\rm MRE}}\right]^{1/21}
   \label{eq:yBDef}
\end{equation}
in terms of the quantity
\begin{equation}
  x_B(t) ~\equiv~ \left[\frac{384}{85} 
    \frac{M_P^6M_3^7\zeta_a^4\zeta_b^7 a_{\rm MRE}^4 \widetilde{x}^{25}}
    {M_1 M_2 (M_1+M_2)^8 (t-t_B)}\right]^{1/21}~.
  \label{eq:xBDef}
\end{equation}
We note that $x_B(t)$ depends on $t$ but not on $x$.  Since the expression for 
$a_B(x)$ in Eq.~(\ref{eq:MNote_aBfull}) is a piecewise function of $x$, the quantity 
$y_B(x,t)$ is likewise a piecewise function of $x$.  Explicitly, we find that
\begin{eqnarray}
  && \!\!\!\!\!\!\rule[-6pt]{0pt}{12pt} y_B(x,t) ~=~ \nonumber \\
  && \!\!\begin{cases}
    \displaystyle x_B\left(\frac{x}{\widetilde{x}}\right)^{37/21}
    & \displaystyle  1 > \frac{x}{\widetilde{x}} > Q_s \\
    \displaystyle x_B\left[\frac{ a_{\rm PBH}}{a_{\rm MRE}Q_{\rm PBH}^{1/\overline{w}}} 
     \right]^{\frac{4}{21}}
    \left(\frac{x}{\widetilde{x}}\right)^{\frac{25+4/\overline{w}}{21}}
    & \displaystyle Q_s > \frac{x}{\widetilde{x}} > Q_{\rm PBH} \\
    \displaystyle x_B\left[\frac{ a_{\rm PBH}}{a_{\rm MRE}} 
     \right]^{\frac{4}{21}}
    \left(\frac{x}{\widetilde{x}}\right)^{\frac{25}{21}}
    & \displaystyle Q_{\rm PBH} > \frac{x}{\widetilde{x}} > Q_f \\
    \displaystyle x_B\left[\frac{a_f}{a_{\rm MRE} Q_f^{1/w_c}}\right]^{\frac{4}{21}}
      \left(\frac{x}{\widetilde{x}}\right)^{\frac{25+4/w_c}{21}} 
    & Q_f > \displaystyle \frac{x}{\widetilde{x}}~. 
   \end{cases}
   \nonumber \\
  \label{eq:yBExplicit}
\end{eqnarray}
While this expression is somewhat complicated, the manner in which $y_B$ scales 
with $x$, on $t$, and the masses $M_1$, $M_2$, and $M_3$ is in fact relatively
straightforward.  Since it is almost always the case that 
$\tau_m(M_1,M_2,M_3,x,y) \gg t_B(M_1,M_2,x)$ for the binary configurations which 
collectively dominate the merger rate a given time $t$,  one can typically
approximate $t - t_B \approx t$ in Eq.~(\ref{eq:xBDef}).  In this approximation, 
we have
\begin{equation}
  y_B(x,t) ~=~ \displaystyle A_B(t)\, x^{(25+\beta)/21}~,
  \label{eq:yBGeneralScalingWithx}
\end{equation}
where the value of $\beta$ and the manner in which the $x$-independent prefactor 
$A_B(t)$ scales with $M_1$, $M_2$, $M_3$, and $t$ both depend on the value of $w$ 
during that epoch.  In particular,
\begin{equation}
  \beta ~=~ \begin{cases}
   \displaystyle 4/w & w ~>~ 0 \\
   \displaystyle 0 & w ~=~ 0~,
    \end{cases}
\end{equation}
while $A_B(t)$ scales with these quantities as
\begin{equation}
  A_B(t) ~\sim~ \frac{M_3^{1/3}}
    {(M_1M_2\,t)^{1/21}(M_1+M_2)^{(24+\beta)/63}}~.
  \label{eq:ABPref}
\end{equation}
As we shall see, this scaling behavior plays a crucial role in determining how
$\Gamma_+(M,t)$ and $\Gamma_-(M,t)$ depend on $M$ and $t$ for any given choice of model 
parameters.

Given these results, we find that the general expression for $\mathcal{R}_3(M_1,M_2,M_3)$ 
in Eq.~(\ref{eq:MNote_MergerRateScriptR}) simplifies in the context of our 
PBH-induced stasis scenario to
\begin{eqnarray}
  &&\!\!\!\!\!\mathcal{R}_3(M_1,M_2,M_3) ~=~ 
    \frac{8\pi^2 \delta_{\rm BH}^2 }{21} \left[\prod_{j=1}^3 a^3f_{\rm BH}(M_j)\right]
      \nonumber \\
  && ~~~~~~\times\,\int_{\min\{
  x_F,\,
  x_p(M_1,M_2)\}}
    ^{x_F} dx\, 
    \Bigg[\mathcal{I}(x)\Theta\big(y_B(x)-x\big) \nonumber \\
  && ~~~~~~\times\,  
    \Theta\big(y_B(x)-x_p(M_1,M_3)\big)\Bigg]~, 
  \label{eq:MNote_MergerRateScriptR3_Simp1}
\end{eqnarray}
where we have once again suppressed the explicit time-dependence in quantities such
as $y_B(x,t)$ and $f_{\rm BH}(M,t)$ in order to avoid notational clutter and where, 
for future reference, we have defined
\begin{equation}
  \mathcal{I}(x) ~\equiv~ \frac{x^2y^3_B(x)}{t-t_B} e^{-N(M_1,y_B(x))}~.
  \label{eq:IofxDef}
\end{equation}

The upper limit of integration $x_F$ in 
Eq.~(\ref{eq:MNote_MergerRateScriptR3_Simp1}) follows from the Heaviside 
function in Eq.~(\ref{eq:DeltFuncAlgebra}), which we can rewrite as a function of
$x$ by inverting the relation between $x$ and $t_B$ in Eq.~(\ref{eq:MNote_tBfull}).  
Given that the decoupling time for binaries with 
$Q_f < x/\widetilde{x} < Q_{\rm PBH}$ is $t_{\rm PBH}$, we have
\begin{equation}
  x_F(t) ~\equiv~ \begin{cases}
    \displaystyle\widetilde{x}\,
    \left(\frac{t}{t_{\rm MRE}}\right)^{1/6} &  t_s < t < t_{\rm MRE} \\
    \displaystyle\widetilde{x}\, Q_{\rm PBH}\left(\frac{t}{t_{\rm PBH}}\right)
      ^{\frac{2\overline{w}}{3(1+\overline{w})}} 
      &  t_{\rm PBH} < t < t_s \\
    \displaystyle\widetilde{x}\, Q_f &  t_f < t < t_{\rm PBH} \\
    \displaystyle\widetilde{x}\, Q_f\left(\frac{t}{t_f}\right)
      ^{\frac{2w_c}{3(1+w_c)}} 
      &  t_i < t < t_f~. \\
  \end{cases}    
\end{equation}
We note that $x_F \leq \widetilde{x}$ for all $t_i < t < t_{\rm MRE}$, and thus the 
upper bound $x < x_F$ is more restrictive than the bound $x < \widetilde{x}$ in all 
cases.  We also note that since the comoving quantity $a^3f_{\rm BH}(M_j)$ is 
time-independent at times $t\lesssim \tau_e(M_j)$
within any region of parameter space within which the effect of mergers and 
accretion is negligible, $\mathcal{R}_3(M_1,M_2,M_3)$ depends on $t$ entirely
through $\mathcal{I}(x)$,  through $x_F(t)$, and through the 
Heaviside functions.

The explicit lower limit of integration in the integral over $x$ in 
Eq.~(\ref{eq:MNote_MergerRateScriptR3_Simp1}) accounts for 
fact that this integral vanishes for combinations of $M_1$ and $M_2$ which 
yield {$x_p(M_1,M_2) > x_F$}.  However, it is 
the additional, implicit thresholds imposed by the Heaviside functions
appearing in the integrand which typically determine the lower limit 
of integration.  As we shall see, these thresholds play a central role in
determining not only how $\mathcal{R}_3(M_1,M_2,M_3)$ depends on $M_1$,
$M_2$, and $M_3$, but also how this differential three-body merger rate 
evolves with $t$ for any particular combination of these masses. 

The expressions for the rates $\Gamma_+(M)$ and $\Gamma_-(M)$ which
follow from this form of $\mathcal{R}_3(M_1,M_2,M_3)$ can be obtained 
via numerical integration.  However, in order to provide physical motivation
for the results we obtain in this way, we first pause to examine in 
greater detail how the expression for $\mathcal{R}_3(M_1,M_2,M_3)$ in 
Eq.~(\ref{eq:MNote_MergerRateScriptR3_Simp1}) depends on the individual masses 
$M_1$, $M_2$, and $M_3$.  Our first step in this direction shall be to define 
\begin{equation}
  \Gamma_3(M_1,M_2,M_3) ~\equiv~ 
    \frac{M_2 M_3\mathcal{R}_3(M_1,M_2,M_3)}{a^3f_{\rm BH}(M_1)}~.
    \label{eq:Gamma3}
\end{equation}
The additional factors present in $\Gamma_3(M_1,M_2,M_3)$ relative to 
$\mathcal{R}_3(M_1,M_2,M_3)$ have been included such that 
\begin{equation}
  \Gamma_3(M_1,M_2,M_3) ~=~ 
    \frac{\partial^2\Gamma_-(M_1)}{\partial\log M_2 \partial\log M_3}~.   
\end{equation}

\begin{figure*}
  \centering
    \includegraphics[width=0.8\linewidth]{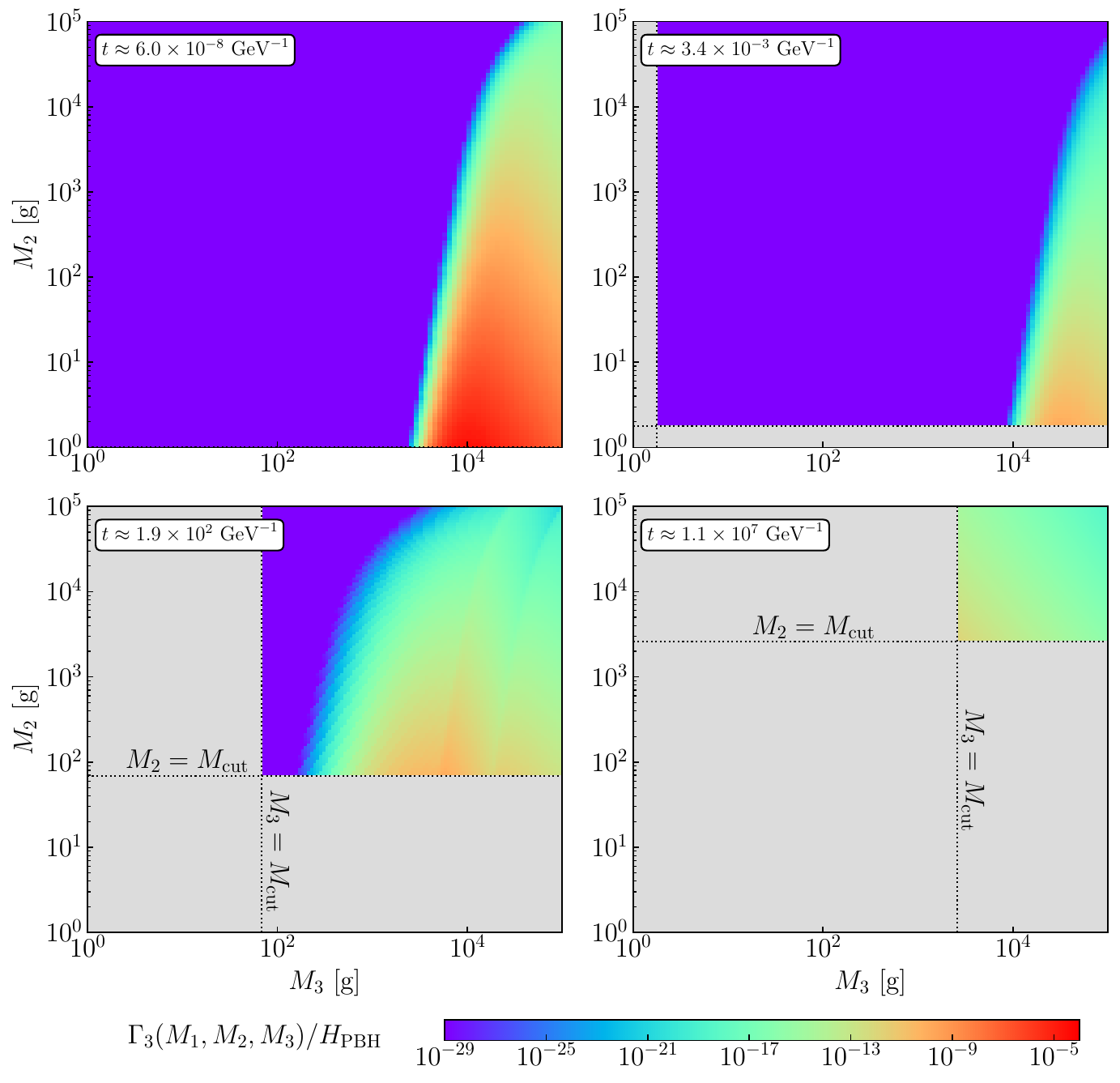}
    \caption{The ratio $\Gamma_3(M_1,M_2,M_3)/H_{\rm PBH}$ of the differential merger 
    rate defined in Eq.~(\ref{eq:Gamma3}) to the value $H_{\rm PBH} = H (t_{\rm PBH})$ of 
    the Hubble expansion rate at the beginning of the stasis epoch 
    for the parameter choices $M_{\rm min} = 1$~g, $M_{\rm max} = 10^5{\rm g}$, 
    $\mathcal{N}_{\rm PBH} = 2$, $\alpha=-2$, and $M = 10^5$~g.  Each panel is
    a density plot of this ratio within the $(M_2,M_3)$-plane evaluated at a different 
    time $t$ within the range $t_i < t < t_s$.  The gray region of each panel indicates 
    the region of the plane wherein either $M_2$ or $M_3$ lies below the corresponding
    value of $M_{\rm cut}(t)$.
  \label{fig:ScanDiffMrgRateMinus}}
\end{figure*}

In Fig.~\ref{fig:ScanDiffMrgRateMinus} we display a series of density plots
which together illustrate how $\Gamma_3(M_1,M_2,M_3)$ evolves over time.  Each of the 
four panels of this figure represents a ``snapshot'' of this quantity, normalized to the
value $H_{\rm PBH} = H (t_{\rm PBH})$ of the Hubble expansion rate at the beginning of the 
stasis epoch, in the $(M_2,M_3)$-plane at a different time between $t_i$ and $t_s$ for a 
population of PBHs characterized by parameter choices $M_{\rm min} = 1$~g, $M_{\rm max} = 10^5$~g, 
$\mathcal{N}_{\rm PBH} = 2$, and $\alpha = -2$ and for $M_1 = 10^5$~g.  The results shown in the 
top left panel correspond to a time during the PBH-formation epoch, those in the top right panel 
correspond to a time shortly after the PBH-domination epoch has concluded, and those in the bottom 
two panels correspond to later times during the stasis epoch.  Within the gray region shown in 
each panel, either $M_2$ or $M_3$ lies below $M_{\rm cut}(t)$ at the 
corresponding time $t$.  This implies that at least one of the PBHs would already have 
evaporated, so no configuration of PBHs with such a combination of masses exists at that time.  

The results shown within each individual panel of Fig.~\ref{fig:ScanDiffMrgRateMinus}
provide a sense of how $\Gamma_3(M_1,M_2,M_3)$ varies with $M_2$ and $M_3$.
Within each panel, we observe that $\Gamma_3(M_1,M_2,M_3)$ decreases monotonically as 
$M_2$ increases, indicating that mergers between a heavy PBH and a very light PBH 
are far more common than mergers between two heavy PBHs.  By contrast, the manner in
which $\Gamma_3(M_1,M_2,M_3)$ depends on $M_3$ is more complicated.  In all but the 
lower right panel of the figure, $\Gamma_3(M_1,M_2,M_3)$ rises, peaks, and then falls 
as $M_3$ increases from $M_{\rm cut}(t)$ to $M_{\rm max}$.  In fact, as we shall see, the 
only reason this behavior is not also manifest in the bottom right panel is that the portion 
of the $(M_2,M_3)$-plane within which $\Gamma_3(M_1,M_2,M_3)$ would otherwise be observed 
to increase with $M_3$ lies within the region wherein $M_3 < M_{\rm cut}(t)$.

The origin of this non-monotonic behavior ultimately stems from an interplay between
the function $\mathcal{I}(x)$ defined in Eq.~(\ref{eq:IofxDef}) and the Heaviside 
functions which modify the limits of integration in the integral over $x$ in  
Eq.~(\ref{eq:MNote_MergerRateScriptR3_Simp1}).  Since Eqs.~(\ref{eq:MNote_NyModel})
and~(\ref{eq:yBGeneralScalingWithx}) together imply that $N(M_1,y_B(x))$ is an 
increasing function of $x$, we may view $\mathcal{I}(x)$ as a function of $x$ which 
exhibits power-law growth for sufficiently small $x$, but is suppressed at large $x$ by 
an exponential cutoff.  The integral in Eq.~(\ref{eq:MNote_MergerRateScriptR3_Simp1})
is in general dominated by the contribution from values of $x$ near the 
value $x_{\rm peak}$ at which this sharply peaked function is maximized --- provided, 
of course, that $x_{\rm peak} \gtrsim \max\{x_p,x_{\rm thresh}\}$, where $x_{\rm thresh}$ 
is the threshold value of $x$ imposed by the Heaviside functions.  

The manner in which both $x_{\rm thresh}$ and $x_{\rm peak}$ scale with $M_2$ and $M_3$ is 
ultimately determined by the manner in which $y_B(x)$ scales with these quantities.  
Using the approximate form of $N(M',y)$ in Eq.~(\ref{eq:NScalingWithyAndt}) 
 and once again approximating $t-t_B \approx t$, we find that 
\begin{eqnarray}
  x_{\rm peak} &~=~& \eta^{7/(25+\beta)}
    \left(A_B B_N^{1/3}\right)^{-21/(25+\beta)} \nonumber \\
  x_{\rm thresh} &~=~& A_B^{-21/(4+\beta)}~, 
  \label{eq:xpeakAndxthresh}
\end{eqnarray} 
where for convenience we have defined $\eta \equiv (39+\beta)/(25+\beta)$.  

Since $B_N(t)$ is independent of both $M_2$ and $M_3$, the relationship between
$x_{\rm peak}$ and $x_{\rm thresh}$ within any individual panel of   
Fig.~\ref{fig:ScanDiffMrgRateMinus} is determined entirely by $A_B(t)$.
Eq.~(\ref{eq:ABPref}) implies that both $x_{\rm peak}$ and $x_{\rm thresh}$ increase 
as $M_2$ increases, but decrease as $M_3$ increases.
However, $x_{\rm thresh}$ is more sensitive to changes in both of these masses.
Thus, for a given value of $t$, it is often the case that $x_{\rm peak} < x_{\rm thresh}$ 
within some regions of the $(M_2,M_3)$-plane, while $x_{\rm peak} > x_{\rm thresh}$ within 
other regions.  

By contrast, we note that $x_p$ --- which, like $x_{\rm thresh}$ provides an upper bound 
on the lower limit of integration in Eq.~(\ref{eq:MNote_MergerRateScriptR3_Simp1}) --- is
completely independent of $M_3$ and is likewise effectively independent of $M_2$ within 
the regime in which $M_2 \ll M_1$.  Within the regime in which $M_2 \gtrsim M_1$, however, 
$x_p$ increases as $M_2$ increases. 

Finally, we note that the value of $\mathcal{I}(x)$ at $x = x_{\rm peak}$ is
\begin{equation}
  \mathcal{I}(x_{\rm peak}) ~=~ \left(\frac{e}{\eta}\right)^{-\eta} \frac{1}{t}
    A_B^{-3(\eta-1)} B_N^{-\eta}~.
\end{equation}
Since $\eta > 1$ for all $-2 \leq \alpha \leq -1$, it it always the case that 
$\mathcal{I}(x_{\rm peak})$ increases as $M_2$ increases, but decreases as 
$M_3$ increases.

These considerations together account for the manner in which $\Gamma_3(M_1,M_2,M_3)$ 
varies throughout the $(M_2,M_3)$-plane within each individual panel of 
Fig.~\ref{fig:ScanDiffMrgRateMinus}.  However, we focus on the results shown in the 
top left panel for purposes of illustration, since the region of the plane within 
which $M_2$ and $M_3$ both exceed $M_{\rm cut}$ is the broadest.  Across the 
entire left edge of the panel, $x_{\rm peak} \ll x_{\rm thresh}$ and $\mathcal{I}(x)$
has negligible support within the integration region.  As a result, $\Gamma_3(M_1,M_2,M_3)$ 
is highly suppressed.  However, as $M_3$ increases with held $M_2$ fixed,
$x_{\rm thresh}$ decreases more rapidly than $x_{\rm peak}$.  Eventually, $x_{\rm thresh}$
decreases to the point at which it overtakes $x_{\rm peak}$ and the peak enters the 
integration region, leading to a dramatic increase in $\Gamma_3(M_1,M_2,M_3)$.  However,
$\mathcal{I}(x_{\rm peak})$ and the additional overall factor 
$M_3 f_{\rm PBH}(M_3)$ in $\Gamma_3(M_1,M_2,M_3)$ also both decrease as $M_3$ increases.  
Thus, as $M_3$ is further increased to the extent where $x_{\rm peak} \gg x_{\rm thresh}$ 
and the entire range of $x$ within which $\mathcal{I}(x)$ has non-negligible support lies 
inside the integration region, this suppression effect eventually causes 
$\Gamma_3(M_1,M_2,M_3)$ to decrease.  

There is a second effect which can also contribute to the decrease in 
$\Gamma_3(M_1,M_2,M_3)$ as $M_3$ increases --- one whose contribution is
frequently far more dramatic than that which results from the decrease in 
$\mathcal{I}(x_{\rm peak})$.  This effect stems from the fact that $x_p$ 
remains fixed as $M_3$ increases, which in turn implies that as $M_3$ increases,
there is a point at which $x_{\rm thresh}$ falls below $x_p$.  Since $x_{\rm peak}$ 
continues to decrease as $M_3$ further increases beyond this point, the range of
$x$ within which $\mathcal{I}(x)$ has significant support eventually also falls below
$x_p$, and $\Gamma_3(M_1,M_2,M_3)$ decreases as a result.  Indeed, as we shall see, 
this effect can have a particularly dramatic effect on the manner in which 
$\Gamma_3(M_1,M_2,M_3)$ varies with $M_3$ in situations in which $\alpha$ is close
to $-1$ and that range of $x$ is particularly narrow.

By contrast, the manner in which $\Gamma_3(M_1,M_2,M_3)$ scales with $M_2$ involves no 
such competition.  As $M_2$ increases, not only does $x_{\rm thresh}$ rise, but   
the product of $\mathcal{I}(x_{\rm peak})$ and the corresponding overall 
factor $M_2 f_{\rm PBH}(M_2)$ in $\Gamma_3(M_1,M_2,M_3)$ actually decreases.  As 
$M_2$ approaches $M_1$, the quantity $M_1+M_2$ becomes sensitive to the value of $M_2$
and $\Gamma_3(M_1,M_2,M_3)$ begins to drop even more steeply with increasing $M_2$.  

The manner in which $\Gamma_3(M_1,M_2,M_3)$ varies across the $(M_2,M_3)$-plane in the 
other panels of Fig.~\ref{fig:ScanDiffMrgRateMinus} exhibits the same qualitative 
behavior.  However, as alluded to above, regions of the plane within which 
certain aspects of this behavior would otherwise be manifest lie within the gray 
region in which $M_3 < M_{\rm cut}$ at late times.  
Indeed, in the bottom right panel, the region of the plane within 
which $\Gamma_3(M_1,M_2,M_3)$ would otherwise be seen to increase with increasing 
$M_3$ lies within this gray region.  Another feature which deserves comment is the 
set of ribbon-like streaks evident in the third panel of Fig.~\ref{fig:ScanDiffMrgRateMinus}.  
These streaks are ultimately a consequence of the discontinuity in $a_B(x)$ which reflects
from the suppression of binary capture during the PBH-dominated epoch.  This 
discontinuity in turn gives rise to discontinuities in $y_B(x)$ and $\mathcal{I}(x)$.
When $x_{\rm thresh}$ lies near one of these discontinuities in $\mathcal{I}(x)$, small 
changes in $M_2$ or $M_3$ can lead to large shifts in $\Gamma_3(M_1,M_2,M_3)$.
The abruptness of these shifts in part reflects the standard approximations discussed above 
Eq.~(\ref{eq:MNote_deoupHflowraw}) which we have invoked in deriving an implicit expression 
for $t_B$.  Indeed, the boundaries of the features which appear in the third panel of 
Fig.~\ref{fig:ScanDiffMrgRateMinus} would presumably appear smoother, given a more detailed 
treatment.  However, we emphasize that whereas binary formation does not completely cease 
during the PBH-dominated epoch, its suppression during this epoch {\it is}\/ physical, and 
the qualitative features in $\Gamma_3(M_1,M_2,M_3)$ to which it leads are physical as well. 

Having examined how $\Gamma_3(M_1,M_2,M_3)$ depends on $M_2$ and $M_3$ for fixed $t$,
we now consider how this quantity evolves with $t$ for fixed $M_2$ and $M_3$. 
A comparison across these different panels reveals that the dependence of 
$\Gamma_3(M_1,M_2,M_3)$ on $t$ is more complicated than its dependence on $M_2$ 
and $M_3$.  This is because $x_{\rm peak}$ and $\mathcal{I}(x_{\rm peak})$ likewise
scale with $t$ in a more complicated manner.  This is primarily due to the 
fact that $B_N(t)$ depends non-trivially on $t$ during the stasis epoch.  During 
the PBH-formation and PBH-dominated epochs, $B_N(t)$ is effectively 
constant and $x_{\rm peak}$ and $x_{\rm thresh}$ both scale with $t$ in exactly the same 
way that they scale with $M_2$.  Thus, $x_{\rm thresh}$ increases with $t$ more 
rapidly than does $x_{\rm peak}$, and the region of the $(M_2,M_3)$-plane within which 
$\Gamma_3(M_1,M_2,M_3)$ is non-negligible consequently decreases.  This effect accounts for 
the difference between the results in the top left panel of 
Fig.~\ref{fig:ScanDiffMrgRateMinus} (which corresponds to a time during
the PBH-formation epoch) and those in the top right panel (which corresponds to a time only 
slightly after the stasis epoch has begun).  By contrast, once stasis begins, the 
situation changes due to the additional time-dependence in $B_N(t)$.  During stasis, 
$x_{\rm peak}$ actually increases {\it more}\/ rapidly than $x_{\rm thresh}$.  As a result,
the region of the $(M_2,M_3)$-plane within which $x_{\rm peak} > x_{\rm thresh}$ expands
over time.  This accounts for the difference between the results shown in the top right 
and bottom left panels.  Of course, as $t$ increases during stasis, $M_{\rm cut}$ also 
increases, and the region of the plane within which there still exist configurations of 
PBHs with a given combination of $M_2$ and $M_3$ shrinks to zero as $t\rightarrow t_s$.

\begin{figure*}
  \centering
    \includegraphics[width=0.48\linewidth]{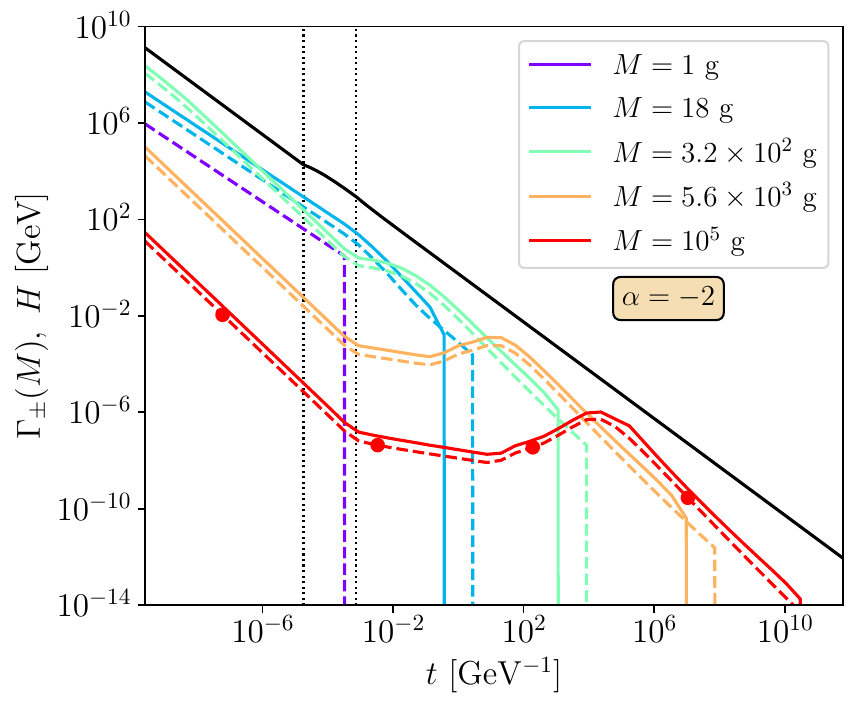}
    \includegraphics[width=0.48\linewidth]{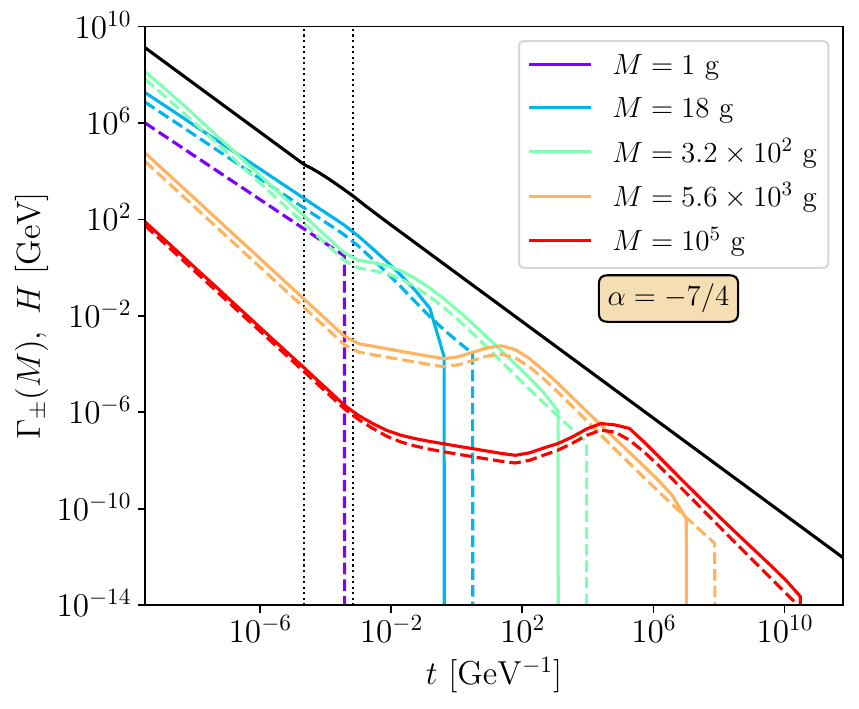}\\
    \includegraphics[width=0.48\linewidth]{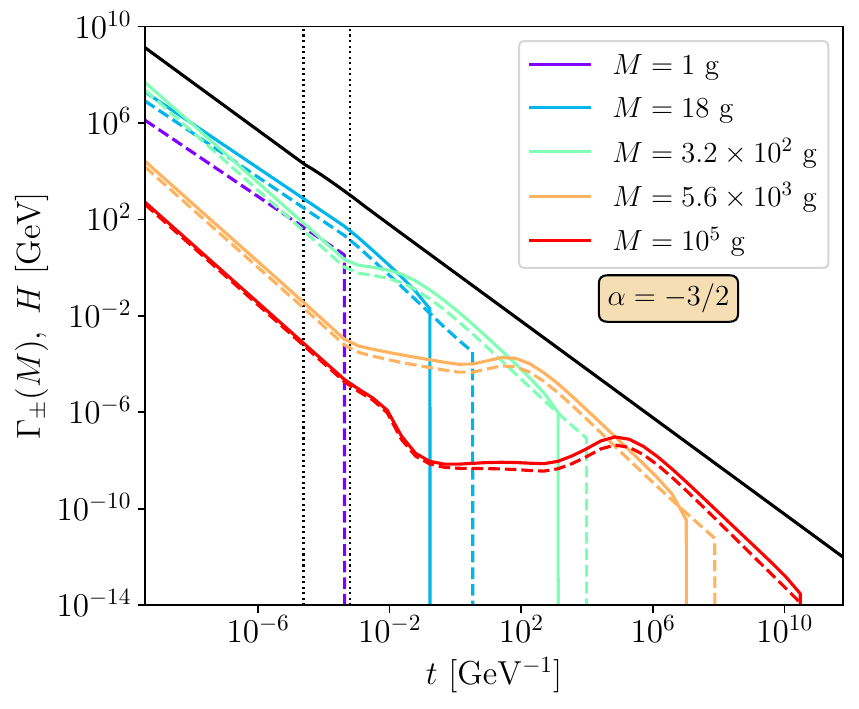}
    \includegraphics[width=0.48\linewidth]{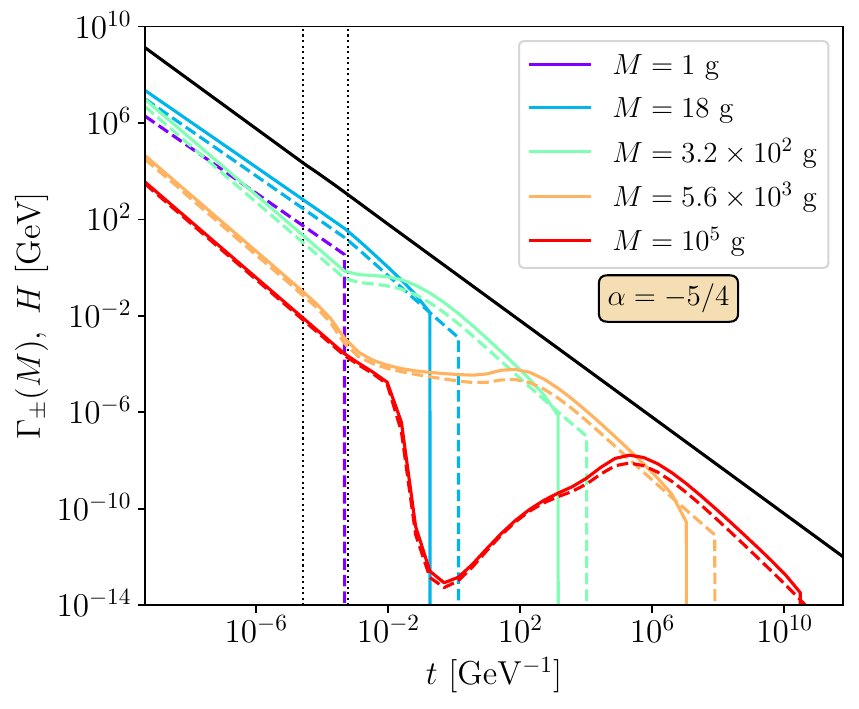}
    \caption{The merger rates $\Gamma_+(M)$ (solid colored curves) and 
    $\Gamma_-(M)$ (corresponding dashed curves), as well as the expansion
    rate $H$, shown as functions of time within the range $t_i < t < t_s$ for a 
    variety of different values of $M$.  The different panels of the figure 
    correspond to different choices of $\alpha$, and the results shown in all 
    panels correspond to the parameter choices $M_{\rm min} = 1$~g, 
    $M_{\rm max} = 10^5$~g, and $\mathcal{N}_{\rm PBH} = 2$.  The four red dots 
    indicated along the $M = 10^5$~g curve in the upper left panel correspond to the 
    four $t$ values for which $\Gamma_3(M_1,M_2,M_3)/H_{\rm PBH}$ is plotted for
    this $M_1$ value in Fig.~\ref{fig:ScanDiffMrgRateMinus}. Since it is always the case 
    that $\Gamma_+(M_{\min}) = 0$, only a dashed curve appears in any of the four panels 
    for $M = 1$~g.
  \label{fig:MrgRates}}
\end{figure*}

In Fig.~\ref{fig:MrgRates}, we show the integrated merger rates $\Gamma_+(M)$  
(solid colored curves) and $\Gamma_-(M)$ (corresponding dashed curves) 
obtained from our expression in Eq.~(\ref{eq:MNote_MergerRateScriptR3_Simp1}) 
for a variety of different values of $M$ as functions of the time $t$.  The value 
of the Hubble parameter (solid black curve) is also plotted alongside these curves 
for reference.  Each panel corresponds to a different choice of $\alpha$, and the 
results shown in all panels correspond to the parameter choices $M_{\rm min} = 1$~g,
$M_{\rm max} = 10^5$~g, and $\mathcal{N}_{\rm PBH} = 2$.  The range of $t$ shown in 
each panel extends from $t_i$ to $t_s$, and the values of $t_f$ and $t_{\rm PBH}$ are 
indicated by the two vertical dashed lines.  The four red dots indicated on the 
$M = 10^5$~g curve in the upper left panel correspond to the values of $t$ for which 
$\Gamma_3(M_1,M_2,M_3)/H_{\rm PBH}$ is plotted for this $M_1$ value in the four panels of 
Fig.~\ref{fig:ScanDiffMrgRateMinus}.  Since it is always the case that 
$\Gamma_+(M_{\min}) = 0$ --- by definition, there exist no lighter PBHs within the
spectrum which could merge to form a PBH of mass $M = M_{\rm min}$ --- only a dashed
curve appears in any of the four panels for $M = 1$~g.

We observe from Fig.~\ref{fig:MrgRates} that the values of $\Gamma_+(M)$ and $\Gamma_-(M)$ 
for any given value of $M$ are typically quite similar in magnitude and evolve in a similar
manner throughout the range of $t$ shown.  The only exceptions are cases in which $M$ lies
very near $M_{\rm max}$.  We also observe that, broadly speaking, $\Gamma_+(M)$ and 
$\Gamma_-(M)$ both decrease over time as the universe evolves from $t_i$ to $t_s$.
At early times, this decrease is primarily due to the phenomenon discussed above, 
wherein the region of the $(M_2,M_3)$-plane for which $x_{\rm peak} > x_{\rm thresh}$
decreases over time.  However, after stasis begins, this region begins to grow again.
As a result, the logarithmic slopes $\partial\Gamma_+(M,t)/\partial \log t$ and 
$\partial \Gamma_-(M,t)/\partial \log t$ both begin to increase.  Indeed, for 
larger values of $M$, this effect persists for a sufficiently long duration that 
$\Gamma_+(M)$ and $\Gamma_-(M)$ both experience a temporary phase of growth.  
However, $\Gamma_+(M)$ and $\Gamma_-(M)$ begin to decrease again once 
the effect of evaporation --- as reflected in the ever-increasing value of
$M_{\rm cut}$ --- begins to impact regions of the $(M_2,M_3)$-plane which 
otherwise would have contributed significantly to $\Gamma_+(M)$ and $\Gamma_-(M)$.  

Interestingly, we also observe that there is an additional effect which becomes
relevant for values of $\alpha$ near the maximum of its allowed range.  Indeed,
within the bottom two panels of Fig.~\ref{fig:MrgRates} --- and particularly 
in the bottom right panel --- we observe that $\partial\Gamma_+(M,t)/\partial \log t$ 
and $\partial \Gamma_-(M,t)/\partial \log t$ initially {\it decrease}\/ during the 
stasis epoch, prior to experiencing the increase which results from the effect
described above.  As a result, a pronounced ``dip'' appears in the $\Gamma_+(M)$ and 
$\Gamma_-(M)$ curves for $M = M_{\rm max}$.

We can understand the emergence of this dip by considering how 
$x_{\rm peak}$, $x_{\rm thresh}$, and $x_p$ behave in the formal $\alpha \to -1$
limit.  After some algebra, one finds that in this limit the expressions in 
Eqs.~(\ref{eq:xpeakAndxthresh}) and~(\ref{eq:xpdef}) reduce to
\begin{eqnarray}
  &&x_{\rm peak} ~\to~ \widetilde{x}
    \left(\frac{2a_s}{a_{\rm MRE}}\right)^{1/3} \nonumber \\
  &&x_{\rm thresh} ~\to~ \widetilde{x}
    \left(\frac{2a_s}{a_{\rm MRE}}\right)^{1/3} \nonumber \\
  && x_p ~\to~ \widetilde{x}
    \left(\frac{2a_s}{a_{\rm MRE}}\right)^{1/3}
    \left(\frac{\max\{M_1,M_2\}}{M_1+M_2}\right)^{1/3}~.~~~~~~~~~
\end{eqnarray}
In other words, $x_{\rm peak}$ and $x_{\rm thresh}$ coincide in this limit, 
while $x_p$ only differs from these other two quantities by a numerical factor 
that ranges between $2^{-1/3}\approx 0.794$ and $1$, depending on the relationship 
between $M_1$ and $M_2$.  We also note that all three of these quantities are 
independent of $M_3$ in this limit.

As we move away from the strict $\alpha \to -1$ limit and consider values of $\alpha$
which are close to but not precisely equal to $-1$, we find that $x_{\rm peak}$, 
$x_{\rm thresh}$, and $x_p$ remain quite similar.  However, $x_{\rm peak}$ and 
$x_{\rm thresh}$ are in general no longer precisely equal and do depend --- though 
not terribly sensitively --- on $M_3$.  Moreover, for such values of $\alpha$, 
the range of $x$ within which $\mathcal{I}(x)$ is non-negligible reduces to a narrow 
spike around $x \approx x_{\rm peak}$.  Thus, if we consider how $\Gamma_3(M_1,M_2,M_3)$ 
depends on $M_3$ for fixed $M_1$ and $M_2$ at any time $t < t_{\rm PBH}$, we find that
this differential merger rate is negligible for small $M_3$ and increases abruptly once 
$M_3$ reaches the critical value at which $x_{\rm thresh}$ drops below $x_{\rm peak}$ 
and this spike enters the integration region.  However, since $x_p$ is also very close to 
$x_{\rm peak}$, only a small additional increase in $M_3$ beyond this critical value 
results in the spike falling below $x_p$ and $\Gamma_3(M_1,M_2,M_3)$ once again becoming 
negligible.  Thus, for $\alpha$ within this regime, the value of $\Gamma_3(M_1,M_2,M_3)$ 
is non-negligible only within a narrow band within the $(M_2,M_3)$-plane. 

The location of this band within the $(M_2,M_3)$-plane depends on the value of $M_1$.
In particular, Eq.~(\ref{eq:ABPref}) implies that for larger $M_1$, the band is located
at smaller values of $M_2$ and larger values of $M_3$.  Moreover, regardless of the 
value of $M_1$, at times $t < t_{\rm PBH}$, the band shifts as $t$ increases toward 
smaller values of $M_2$ and larger values of $M_3$.  Thus, for large $M_1 \sim M_{\rm max}$, 
it migrates over time into the corner of the $(M_2,M_3)$-plane where $M_2\sim M_{\rm min}$ 
and $M_3 \sim M_{\rm max}$ --- the region of the plane wherein the coalescence time $\tau_m$  
is maximized.  While this migration is a universal behavior which occurs for all $\alpha$, 
the region of the $(M_2,M_3)$-plane within which $\Gamma_3(M_1,M_2,M_3)$ is non-negligible is 
significantly broader for lower values of $\alpha$.

Once stasis begins, however, the PBHs with $M \sim M_{\rm min}$ rapidly evaporate, 
and the binary configurations for which $\Gamma_3(M_1,M_2,M_3)$ is sizable vanish 
long before they have the opportunity to coalesce.  As a result 
$\Gamma_+(M,t)$ and $\Gamma_-(M,t)$ drop precipitously.  Eventually, as $t$ further 
increases and the region of the $(M_2,M_3)$-plane within which $\Gamma_3(M_1,M_2,M_3)$ 
is sizable broadens, these rates begin to grow again and eventually become 
comparable to the corresponding rates obtained for smaller values of $\alpha$.

By far the most important message of Fig.~\ref{fig:MrgRates}, however, is that in all four
panels of the figure, $\Gamma_+(M)$ and $\Gamma_-(M)$ both remain well below $H$ for all 
$M_{\rm min} < M < M_{\rm max}$ throughout the entire period from $t_i < t < t_s$.  
The effect of mergers on $f_{\rm BH}(M,t)$ is therefore negligible in comparison with the 
effect of expansion throughout this period.  We therefore conclude that the shape of 
$f_{\rm BH}(M,t)$ is not significantly distorted by mergers either prior to or during
the stasis epoch.  In other words, we conclude that $\MD = M_{\rm max}$.  
Moreover, while the results shown in Fig.~\ref{fig:MrgRates} reflect 
the particular choices of $M_{\rm min}$, $M_{\rm max}$, and $\mathcal{N}_{\rm PBH}$ that
we have adopted for purposes of illustration therein, we have verified that the same conclusions 
hold for essentially any other viable combination of these model parameters as well.  Thus,
we conclude that mergers between PBHs have only a negligible impact on the dynamics which
give rise to stasis in our PBH-induced stasis model and can therefore be ignored.


\section{Accretion\label{sec:accretion}}


In general, the contribution $(dM/dt)_{\rm acc}$ to the rate at which the mass 
$M$ of a given PBH changes as a result of accretion within the early universe 
depends on $M$ itself, on the cosmological expansion rate, and on the nature 
of the material being accreted.  The effect of accretion on PBH masses has been 
investigated extensively within the context of the thermal expansion history
associated with the standard cosmology~\cite{Ricotti:2007jk,Ricotti:2007au,
Rice:2017avg,DeLuca:2020bjf,DeLuca:2020fpg,DeLuca:2020qqa}.  In this section, 
we examine its effect within the context of our PBH-induced stasis cosmology.

Within the regime in which the Schwarzschild radius of the PBH is much smaller than the 
Hubble horizon, $(dM/dt)_{\rm acc}$ may be estimated by means of a quasi-stationary
approximation~\cite{Bondi:1952ni,Zeldovich:1967lct}.  In this approximation, the effect 
of cosmological expansion is neglected except for its impact on the energy density of the 
universe, which is still taken to be approximately equal to $\rho_{\rm crit}$ at large 
distances away from the PBH.  Applying this quasi-stationary approximation to the case of 
an isolated Schwarzschild black hole of mass $M$ immersed in a perfect fluid with 
cosmological abundance $\Omega_F$ and equation-of-state parameter $0 < w_F < 1$, one may 
derive an expression for $(dM/dt)_{\rm acc}$, which takes 
the form~\cite{Babichev:2004yx,Babichev:2005py} 
\begin{equation}
  \left(\frac{dM}{dt}\right)_{\rm acc} ~=~ 
    \frac{4\pi A(w_F) M^2 (1+w_F) \Omega_F \rho_{\rm crit}}{M_P^4}~, 
  \label{eq:AccretRateGen}
\end{equation}
where $A(w_F)$ is a numerical factor which depends on $w_F$.  Formally, for 
a fluid with a constant value of $0 < w_F < 1$, one finds that this numerical
factor is given by~\cite{Babichev:2005py}
\begin{equation}
  A(w_F) ~=~ \frac{(1+3w_F)^{(1+3w_F)/(2w_F)}}{4w_F^{3/2}}~,
  \label{eq:AofwF}
\end{equation}
which has the property that $A(w_F) > 4$ for all $w_F$ within this range.  
However, this expression diverges in the $w_F \rightarrow 0$ limit and 
is valid only within the regime in which $w_F$ differs significantly from zero.

The effect of accretion in the opposite regime has been examined
by a number of authors~\cite{Bertschinger:1985pd,deJong:2021bbo,DeLuca:2021pls} ---
typically in the context in which the fluid with $w_F \approx 0$ is a gas of non-relativistic 
particles which represents a significant fraction of the total energy density of the 
universe.  The results of these studies 
suggest that the effect of accretion on PBH masses can indeed be substantial in
this regime, though there exist significant uncertainties as to its quantitative impact.
Thus, while it is likely that the accretion of a fluid with $w_F \approx 0$ and 
$\Omega_F \sim \mathcal{O}(1)$ onto a population of PBHs for a protracted period 
would induce distortions in the shape of $f_{\rm BH}(M,t)$, the precise extent to 
which this effect constrains the emergence or duration of stasis within the parameter 
space of our model is unclear.  

Within the context of this model, the only fluid with 
a non-negligible abundance which can accrete onto the PBHs and which can potentially
have an equation-of-state parameter near zero is the perfect fluid which dominates
the energy density of the universe during the PBH-formation epoch, which has 
$w_c \approx 0$ when $\alpha$ is near the maximum of its allowed range.  
In what follows, we shall therefore focus primarily on the regime wherein $\alpha$ is 
not too close to this maximum value and $(dM/dt)_{\rm acc}$ can reliably be modeled 
using the expression in Eq.~(\ref{eq:AccretRateGen}).  More specifically, we focus on 
the regime in which $\alpha < -1.2$, for which $w_c > 1/9$.

During the PBH-formation epoch, the material accreted by the PBHs 
is the perfect fluid which dominates the energy density of the universe.
Thus, during this epoch, $\Omega_F = 1$ and $w_F = w_c$.  
However, since the Schwarzschild radius of a PBH is comparable to the 
size of the Hubble horizon at the time it forms due to gravitational 
collapse, cosmological expansion has a significant effect on the accretion 
rate~\cite{Carr:1974nx,Carr:2010wk}.  The quasi-stationary approximation on which 
Eq.~(\ref{eq:AccretRateGen}) is predicated is therefore unreliable for a PBH of 
initial mass $M_i$ until $H^{-1} \gg 2GM_i$ --- or, equivalently, until 
$t \gg t_p(M_i)$.  In what follows, for concreteness, we shall assume that this 
approximation can safely be invoked for such a PBH at times 
$t > \theta_{\rm QS} t_p(M_i)$, where $\theta_{\rm QS} \gg 1$ is independent 
of $M_i$.

There exists some uncertainty regarding the manner in which a PBH of initial mass 
$M_i$ evolves from $t_p(M_i)$ until the time at which the quasi-stationary approximation 
becomes valid.  Nevertheless, it has been 
shown~\cite{Carr:1974nx,BicknellGenSS,Custodio:1998pv,Harada:2004pf,Harada:2004pe} 
that the sustained period of rapid, self-similar growth which would appear to follow
from Eq.~(\ref{eq:AccretRateGen}) for a PBH with a radius comparable to the 
Hubble horizon does not in fact occur --- at least in situations in which $w_F$ 
differs significantly from zero.  Moreover, for such values of $w_F$, it is commonly 
assumed that $(dM/dt)_{\rm acc}$ is small and that its effect on $M$ is not terribly 
significant~\cite{Carr:1975qj}.  In what follows, we shall therefore assume that the 
mass of a PBH is not significantly modified from $M_i$ while the Hubble horizon remains 
comparable to its Schwarzschild radius.  We shall consider the implications of relaxing 
this assumption in Sect.~\ref{sec:conclusions}.

By contrast, once $t \gg t_p(M_i)$ and $(dM/dt)_{\rm acc}$ can reliably be 
approximated using Eq.~(\ref{eq:AccretRateGen}), evaluating the effect of accretion
on the PBH mass spectrum is far more straightforward.  Indeed, this equation implies
that
\begin{equation}
  \frac{dM}{dt} ~=~ 
    \left[\frac{2A_F(w_c)}{3(1+w_c) M_P^2}\right] 
    \frac{M^2}{t^2}
  \label{eq:AccretRateGenSimpStasis}
\end{equation}
during the remainder of the PBH-formation epoch.  Solving this equation, we find 
that the mass of a PBH during the stasis epoch is given at times 
$t_p(M_i) \ll  t < t_f$ by
\begin{equation}
  M(t) ~=~ M_i 
    \left[1 - C_f(w_c)
    \left(1-\frac{\theta_{\rm QS}t_p(M_i)}{t} \right) \right]^{-1}~,
  \label{eq:AccretionMoftinStasis}
\end{equation}
where we have defined
\begin{equation}
  C_f(w_c) ~\equiv~ \frac{2A_F(w_c)M_i}{3M_P^2(1+w_c)\theta_{\rm QS}t_p(M_i)} 
    ~=~ \frac{\gamma A_F(w_c)}{2\theta_{\rm QS}}~.
  \label{eq:Cfofwc}
\end{equation}

Since $C_f(w_c)$ is independent of $M_i$, it follows that
within the regime in which $t_f \gg \theta_{\rm QS} t_i$ --- the regime in which  
the quasi-static approximation becomes valid for our entire population of PBHs
well before the PBH-formation epoch ends --- the mass of every PBH at $t = t_f$
is rescaled relative to its corresponding initial mass by an effectively universal 
factor.  In particular, taking $\gamma = 1$, we have
\begin{equation}
  M(M_i,t_f) ~\approx~ M_i \left[1-\frac{A_F(w_c)}{2\theta_{\rm QS}}\right]^{-1}~.
  \label{eq:AccretionPBHFormEpochMScaling}
\end{equation}
Within this regime, then, the PBH mass spectrum at $t = t_f$ is   
\begin{equation}
  f_{\rm BH}(M,t_f) ~=~ \left(\frac{a}{a_i}\right)^{-3} 
    \left[1-\frac{A_F(w_c)}{2\theta_{\rm QS}}\right]^{1-\alpha} f_{\rm BH}(M_i,t_i)~.    
\end{equation}
In other words, the overall normalization of $f_{\rm BH}(M,t)$ is increased
relative to what it otherwise would have been in the absence of accretion, and the
range of masses within the spectrum is shifted upward, but the overall power-law scaling 
behavior of $f_{\rm BH}(M,t)$ with $M$ --- and even the value of the power-law 
exponent --- remains unchanged.  Moreover, for $\alpha < -1.2$, we find that the proportional 
increase in the the mass of each PBH during the period from $t_p(M_i)$ to $t_f$ is bounded 
from above by $[M(M_i, t_f) - M_i]/M_i \lesssim 0.33$.  Thus, we conclude that the effect of 
accretion on the range of PBH masses during the PBH-formation epoch is not terribly 
significant for values of $\alpha$ within this regime.

Given that the effect of accretion on $f_{\rm BH}(M,t)$ within this regime is simply an 
overall rescaling accompanied by a shift in the range of $M$ for which this function
receives non-vanishing support, we note that this effect can be compensated by a 
redefinition of $M_{\rm min}$, $M_{\rm max}$, and $\mathcal{N}_{\rm PBH}$.  In other 
words, the apparent values of these parameters at $t = t_f$ are different --- and all 
slightly larger --- than they would have been in the absence of accretion.  However,
since these shifts are indeed quite slight, their impact on the results of the 
merger-rate analysis we performed in Sect.~\ref{sec:results}, as well as on other 
aspects of our PBH-induced stasis cosmology, is insignificant.
Throughout the remainder of this section, we shall account for the effect of accretion 
within the regime in which $t_f \gg \theta_{\rm QS} t_i$ via such a redefinition of
$M_{\rm min}$, $M_{\rm max}$, and $\mathcal{N}_{\rm PBH}$.  In other words, we shall
redefine $M_i$ such that $M_i \rightarrow M(M_i,t_f)$. 
 
By contrast, within the regime in which $t_f \sim \theta_{\rm QS}t_i$, the factor by
which $M$ increases relative to $M_i$ between $t_p(M_i)$ and $t_f$ is not independent of
$M_i$.  Within this regime, distortions to $f_{\rm BH}(M,t)$ can arise and the 
effect of accretion cannot be accounted for via a redefinition of $M_{\rm min}$, 
$M_{\rm max}$, and $\mathcal{N}_{\rm PBH}$.  Thus, such distortions {\it can}\/
meaningfully impact the dynamics which give rise to stasis.  

Nevertheless, there are reasons why 
the impact of these distortions on the expansion history is not expected to be 
particularly severe.  First of all, they arise only at the upper end of the PBH mass 
spectrum.  Indeed, since the collapse time scales with $M_i$ as $t_p(M_i) \sim M_i$, 
the distortions in the shape of $f_{\rm BH}(M_i,t_f)$ which arise due to accretion are 
only meaningful for $M_i \sim M_{\rm max}$.  By contrast, for PBHs with $M_i \ll M_{\rm max}$ 
the collapse time is such that $t_f \gg \theta_{\rm QS} t_p(M_i)$.  The factor by 
which $M$ increases relative to $M_i$ for these lighter PBHs is therefore effectively 
independent of $M_i$ and given by Eq.~(\ref{eq:AccretionPBHFormEpochMScaling}).  
This implies that distortions to the PBH mass spectrum which develop during the 
PBH-formation epoch do not disrupt stasis outright, but at worst result in the
stasis epoch ending prematurely by a few $e$-folds, at the point when the evaporation 
of the heavier PBHs would have become important for sustaining stasis.   
Moreover, as discussed above, even when distortions to $f_{\rm BH}(M,t)$ do arise, 
the magnitude of the distortions is at most at the $33\%$ level.  Thus, the impact 
of these distortions on the matter abundance $\Omega_{\rm BH}$ as the system begins 
to depart from stasis is far from dramatic.

One criterion which can be used to quantify the extent of these distortions is the 
ratio of the factor $M(M_{\rm min},t_f)/M_{\rm min}$ by which the mass of a PBH
with $M_i = M_{\rm min}$ grows during the PBH-formation epoch to the 
factor $M(M_{\rm max},t_f)/M_{\rm max}$ by which the mass of a PBH with 
$M_i = M_{\rm max}$ grows during this epoch.  In particular, the impact of these 
distortions on stasis should be negligible whenever the criterion $J \ll 1$ 
is satisfied, where we have defined
\begin{equation}
  J ~\equiv~ \frac{M(M_{\rm min},t_f)M_{\rm max}}{M(M_{\rm max},t_f)M_{\rm min}} -1~.
  \label{eq:MDistortRatio}
\end{equation}
By contrast, when this criterion is not satisfied, the stasis epoch ends prematurely,
as discussed above, but still can be of significant duration.

As the universe enters the PBH-dominated epoch, the contribution to energy density 
of the universe from the perfect fluid which dominated that energy density
during the PBH-formation epoch becomes subleading.  Nevertheless, PBHs can still 
accrete material from this fluid during the PBH-dominated epoch.  Thus, once again,
we have $w_F = w_c$.  However, the abundance $\Omega_F(t)$ during 
the PBH-dominated epoch evolves with time according to the relation
\begin{equation}
  \Omega_F(t) ~\approx~ \frac{1}{2}\left(\frac{t}{t_f}\right)^{-2w_c}~,  
\end{equation} 
where we have taken $\Omega_F(t_f) = 1/2$ at the transition point between the 
two epochs.  As a result, Eq.~(\ref{eq:AccretRateGen}) reduces to
\begin{equation}
  \left(\frac{dM}{dt}\right)_{\rm acc} ~=~ 
    \left[\frac{(1+w_c)A(w_c)t_f^{2w_c}}{3M_P^2}\right] \frac{M^2}{t^{2(1+w_c)}}~.
\end{equation}
during this epoch.  Solving this differential equation for $M$, we find that
\begin{equation}
  M(t) ~=~ M(t_f)\left[1 - C_{\rm PBH}(w_c)
      \left(1-\frac{t_f^{1+2w_c}}{t^{1+2w_c}}\right)\right]^{-1}~,
  \label{eq:AccretRateGenSimpPBHDom}
\end{equation}
where we have defined
\begin{equation}
  C_{\rm PBH}(w_c) ~\equiv~ \frac{(1+w_c)A(w_c)M(t_f)}{3(1+2w_c)M_P^2\,t_f}~.
  \label{eq:AccCoeffPBHDom}
\end{equation}
We note that $C_{\rm PBH}(w_c)$, unlike $C_f(w_c)$, depends non-trivially on $M_i$.
As a result, accretion during this epoch leads to distortions in the shape of 
$f_{\rm BH}(M,t)$ across the entire PBH mass spectrum.

Since we compensate for the effect of accretion during the PBH-formation epoch 
via a redefinition of $M_i$, as discussed above, we take $M(t_f) = M_i$ in 
Eqs.~(\ref{eq:AccretRateGenSimpPBHDom}) and~(\ref{eq:AccCoeffPBHDom}).
We also note that in deriving Eq.~(\ref{eq:AccretRateGenSimpPBHDom}), we have 
assumed that the quasi-stationary approximation is valid at all times 
$t_f < t < t_{\rm PBH}$, and thus that $2 M_{\rm max} G H(t_f) \ll 1$.

Finally, during the stasis epoch, the surrounding fluid is the radiation 
into which the PBHs decay, so we have $\Omega_F = \barOmega_\gamma$ and 
$w_F = 1/3$.  Since the value of $w = \overline{w}$ for the universe as a whole is also 
effectively constant during this epoch, we have 
\begin{equation}
  \rho_{\rm crit} ~=~ \frac{M_P^2}{6\pi(1+\overline{w})^2t^2}~.  
\end{equation}
As discussed below Eq.~(\ref{eq:MNote_timePBHevap}), most of the mass lost by a PBH
of mass $M$ as it evaporates is lost at times $ t \sim \tau_e(M)$.  Thus to a good
approximation, we may ignore the effect of evaporation on the evolution of $M$ at 
times $t \ll \tau_e(M)$.  Thus, during the stasis epoch, $(dM/dt)_{\rm acc}$ represents 
the only potentially non-negligible contribution to the overall rate of change of $M$. 
Since the quasi-stationary approximation on which Eq.~(\ref{eq:AccretRateGen}) is 
predicated is indeed valid during this epoch, we have
\begin{equation}
  \frac{dM}{dt} ~=~ 
    \left[\frac{16\sqrt{3} \,\overline{w} }{(1+\overline{w})^2 M_P^2}\right] 
    \frac{M^2}{t^2}~.
  \label{eq:AccretRateGenSimpStasis}
\end{equation}
Solving this equation, we find that the mass of a PBH during the stasis epoch 
is given at times $t_{\rm PBH} \lesssim  t \ll \tau_p(M)$ by
\begin{equation}
  M(t) ~=~ M(t_{\rm PBH}) 
    \left[1 - C_s(\overline{w})
    \left(1-\frac{t_{\rm PBH}}{t} \right) \right]^{-1}~,
  \label{eq:AccretionMoftinStasis}
\end{equation}
where the mass $M(t_{\rm PBH})$ of the PBH at the beginning of the 
stasis epoch is given by taking $t=t_{\rm PBH}$ in Eq.~(\ref{eq:AccretRateGenSimpPBHDom}) 
and where we have defined
\begin{equation}
  C_s(\overline{w}) ~\equiv~ \frac{16\sqrt{3} \,\overline{w}M(t_{\rm PBH}) }
    {(1+\overline{w})^2M_P^2 t_{\rm PBH}}~.
  \label{eq:AccCoeffStasis}
\end{equation}

In order to assess the net effect of accretion on the mass of an individual 
PBH during the PBH-domination and stasis epochs, we consider the proportional 
increase in mass
\begin{equation}
  \frac{\Delta M}{M} ~\equiv~ \frac{M(M_i,\tau_e(M_i)) - M_i}{M_i}~    
\end{equation}
that such a PBH experiences during the period from $t_f$ until the time $\tau_e(M_i)$ 
at which it would have completely evaporated in the absence of accretion.  

Since the evaporation time of a PBH increases with $M$ in accord with 
Eq.~(\ref{eq:MNote_timePBHevap}), a PBH which would have evaporated at
$t = \tau_e(M_i)$ in the absence of accretion would still exist at this time
if it accretes material at any time after it was produced.  Moreover,
Eqs.~(\ref{eq:AccretRateGenSimpPBHDom}) and~(\ref{eq:AccretRateGenSimpPBHDom})
indicate that $\Delta M/M$ increases with $M$ across any particular population
of PBHs.  This implies that if the value $(\Delta M/M)_{\rm max}$ obtained by taking 
$M_i = M_{\rm max}$ within such a population is negligible, the effect of accretion
is negligible across the entire PBH mass spectrum.  Given this, we focus on evaluating 
$(\Delta M/M)_{\rm max}$ in what follows. 

In order to determine in which regions of the parameter space of our PBH-induced 
stasis model the criterion $(\Delta M/M)_{\rm max} \ll 1$ is satisfied, we survey that 
parameter space numerically, using the full expressions in Eqs.~(\ref{eq:AccretRateGenSimpPBHDom})
and~(\ref{eq:AccretionMoftinStasis}).  However, we also note that a reasonably 
simple analytic approximation for this condition can be formulated in a straightforward 
manner.  Indeed, within the regime in which $\Delta M/M \ll 1$, the second term within 
the square brackets in each of Eqs.~(\ref{eq:AccretRateGenSimpPBHDom}) 
and~(\ref{eq:AccretionMoftinStasis}) is much less than unity.  Expanding 
$\Delta M/M$ and retaining terms up to linear order in each of these small
dimensionless quantities and taking $M_i = M_{\rm max}$ yields
\begin{eqnarray}
  \left(\frac{\Delta M}{M}\right)_{\rm max} &\,\approx\,& M_{\rm max} \Bigg\{
    C_{\rm PBH}(w_c)
      \left[1-\left(\frac{t_f}{t_{\rm PBH}}\right)^{1+2w_c}\right] \nonumber \\
    && ~~~~~~+\, C_s(\overline{w})
    \left(1-\frac{t_{\rm PBH}}{\tau_e(M_i)} \right) \Bigg\}
\end{eqnarray}
at $t = \tau_e(M_{\rm max})$.  Making use of 
Eqs.~(\ref{eq:AccCoeffPBHDom}) and~(\ref{eq:AccCoeffStasis}), along with the
fact that
\begin{equation}
   \frac{t_{\rm PBH}}{t_f} ~=~ \left(\frac{a_f}{a_{\rm PBH}}\right)^{3/2} 
     = e^{-3\mathcal{N}_{\rm PBH}/2}~, 
\end{equation}
in order to simplify this result, we find that the condition under which
$\Delta M/M \ll 1$ for all $M_{\rm min} \leq M \leq M_{\rm max}$ can be
expressed as
\begin{eqnarray}
  \frac{M_{\rm min}^3}{M_{\rm max}} &\,\gg\,& \epsilon M_P^2
  \Bigg[\frac{48\sqrt{3}\overline{w}}{(1+\overline{w})^2}
    \Bigg(1 - \frac{M_{\rm min}^3}{M_{\rm max}^3}\Bigg) \nonumber \\
   & &+\, \frac{(1+w_c)A(w_c)}{1+2w_c}
    \Big(e^{3\mathcal{N}_{\rm PBH}/2}-e^{-3w_c\mathcal{N}_{\rm PBH}}\Big) 
    \Bigg]~. \nonumber \\
  \label{eq:GeneralDeltaMOverMCondit}
\end{eqnarray}

Within the regime in which $M_{\rm min} \ll M_{\rm max}$ and 
$\mathcal{N}_{\rm PBH} \gtrsim 1$, this relation reduces to
\begin{equation}
  \frac{M_{\rm min}^3}{M_{\rm max}} ~\gg~ 
  \left[\frac{48\sqrt{3}\overline{w}}{(1+\overline{w})^2} 
   + \frac{(1+w_c)A(w_c)}{1+2w_c}e^{3\mathcal{N}_{\rm PBH}/2}\right]
    \epsilon M_P^2~.
  \label{eq:SimplifiedDeltaMOverMCondit}
\end{equation}
Indeed, since $M_{\rm min} \ll M_{\rm max}$ within our region of interest
for stasis in this scenario, this more compact formulation of the 
$(\Delta M/M)_{\rm mmax} \ll 1$ condition is valid throughout the vast majority of our
parameter-space region of interest.  Qualitatively, it is clear 
from Eq.~(\ref{eq:SimplifiedDeltaMOverMCondit}) that the regions of 
parameter space within which the effect of accretion on $f(M,t)$  
is the most significant are those in which $M_{\rm min}$ is near its 
lower limit, $M_{\rm max}$ is near its upper limit, and $\mathcal{N}_{\rm PBH}$ 
is large.  It is also apparent from Eq.~(\ref{eq:SimplifiedDeltaMOverMCondit})
that the corresponding constraint contour is not particularly sensitive to the 
value of $\alpha$, except within the regime in which $\alpha$ is very close to 
$-1$ and $A(w_c)$ is therefore extremely large.  However, as we have discussed
above, the expression for $A(w_c)$ in Eq.~(\ref{eq:AofwF}) is likely unreliable in
this regime, and in what follows, we restrict our consideration to values of
$\alpha$ within the range $-2 < \alpha < -1.2$.

\begin{figure*}
  \centering
  \includegraphics[width=0.32\textwidth]{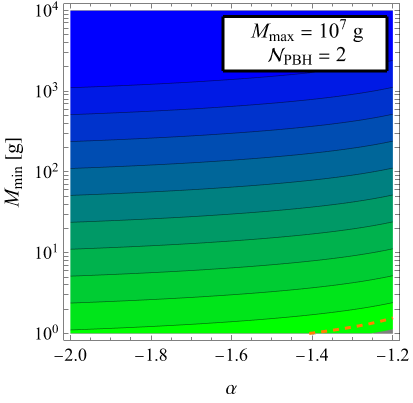}~
  \includegraphics[width=0.32\textwidth]{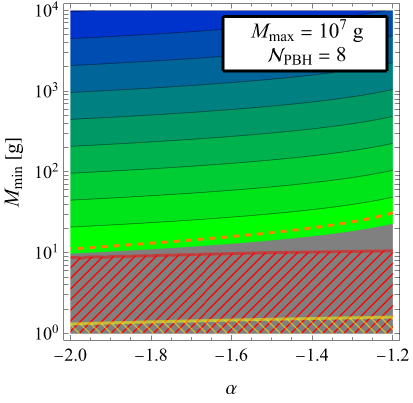}~
  \includegraphics[width=0.32\textwidth]{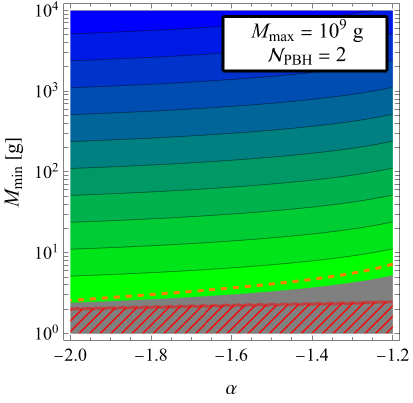}\\
  \includegraphics[width=0.5\textwidth]{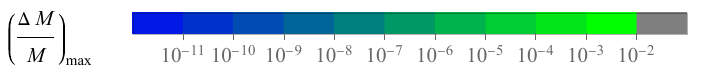}
  \caption{Contours of $(\Delta M/M)_{\rm max}$ within the $(\alpha,M_{\rm min})$-plane at 
    the time $t = \tau_e(M_{\rm max})$ at which a PBH with $M_i = M_{\rm max}$ would have 
    evaporated completely in our stasis cosmology in the absence of accretion --- \ie, at 
    the time at which the stasis epoch would have ended.  The left panel shows the results 
    for $M_{\rm max} = 10^7$~g and $\mathcal{N}_{\rm PBH} = 2$, the middle panel shows the 
    results for $M_{\rm max} = 10^7$~g and $\mathcal{N}_{\rm PBH} = 8$, and the right panel 
    shows the results for $M_{\rm max} = 10^9$~g and $\mathcal{N}_{\rm PBH} = 2$.  
    Within the gray region at the bottom of each panel, $(\Delta M/M)_{\rm max} > 0.01$, 
    indicating a non-negligible change in $M$ for at least the heaviest PBHs.  The red hatched 
    region in each panel indicates the region within which PBHs with $M_i = M_{\rm max}$ form
    sufficiently close to the end of the PBH-formation epoch that the quasi-stationary 
    approximation underlying Eq.~(\ref{eq:AccretRateGen}) may not be reliable.  The 
    yellow hatched region at the bottom of the middle panel indicates the region of the 
    $(\alpha,M_{\rm min})$-plane within which the consistency condition 
    $\mathcal{N}_{\rm PBH} < \mathcal{N}_{\rm PBH}^{({\rm max})}$ is violated.
    Throughout the region above the orange dashed contour in each panel, the 
    condition $J < 0.01$ is satisfied for $\theta_{\rm QS} = 100$.
  \label{fig:DeltaMOverMPlot}}
\end{figure*}

We now present the results of our numerical analysis.  In Fig.~\ref{fig:DeltaMOverMPlot}, 
we show contours of $(\Delta M/M)_{\rm max}$ within the $(\alpha,M_{\rm min})$-plane at the 
time $t = \tau_e(M_{\rm max})$.  The left panel shows the results for
$M_{\rm max} = 10^7$~g and $\mathcal{N}_{\rm PBH} = 2$, the middle panel shows the 
results for $M_{\rm max} = 10^7$~g and $\mathcal{N}_{\rm PBH} = 8$, and the right panel 
shows the results for $M_{\rm max} = 10^9$~g and $\mathcal{N}_{\rm PBH} = 2$. 
Within the gray region at the bottom of each figure, $(\Delta M/M)_{\rm max} > 0.01$, indicating
a non-negligible change in $M$ for at least the heaviest PBHs.  The red hatched region in each 
panel indicates the portion of the plane within which PBHs with $M_i = M_{\rm max}$ form sufficiently 
close to the end of the PBH-formation epoch that their Schwarzschild radii are still comparable 
in size to the Hubble horizon and the quasi-stationary approximation underlying 
Eq.~(\ref{eq:AccretRateGen}) is therefore unreliable.  More precisely speaking,  
this region corresponds to the portion of the plane wherein $2 M G H(t_f) > 0.01$.
The yellow hatched region in the middle panel indicates the region wherein the consistency 
condition $\mathcal{N}_{\rm PBH} < \mathcal{N}_{\rm PBH}^{({\rm max})}$ is violated.
However, we note that since this constraint represents that requirement that
PBHs with $M_i = M_{\rm max}$ are produced before $t_f$, it is in a sense simply a 
weaker version of the constraint indicated by the red hatched region.  
Throughout the region above the orange dashed contour in each panel, the 
condition $J < 0.01$ is satisfied for $\theta_{\rm QS} = 100$.  Below this contour, 
distortions develop near the upper end of the mass spectrum during the PBH-formation epoch
and stasis concludes a few $e$-folds earlier than it otherwise would have.

We observe that there exists a significant region of parameter space within each panel 
of Fig.~\ref{fig:DeltaMOverMPlot} wherein $(\Delta M/M)_{\rm max} < 0.01$ and $J < 0.01$,
while the other consistency conditions we have imposed are likewise satisfied.  
The effect of accretion on the spectrum of PBHs throughout this entire region is 
negligible.  We also observe that $(\Delta M/M)_{\rm max}$ decreases substantially as 
$M_{\rm min}$ increases, but that this quantity is not terribly sensitive to the value 
of $\alpha$ --- provided of course that $\alpha$ is not too close to $-1$ and
the expression in Eq.~(\ref{eq:AccretRateGen}) therefore remains valid.  Comparing the 
results across the three panels of the figure, we also observe that
our negligibility condition $(\Delta M/M)_{\rm max} < 0.01$ becomes more constraining within
the $(\alpha,M_1)$-plane as either $M_{\rm max}$ or $\mathcal{N}_{\rm PBH}$ increases.

Taken together, the results shown in Fig.~\ref{fig:DeltaMOverMPlot} indicate that unless
$\mathcal{N}_{\rm PBH}$ is quite large, $M_{\rm min}$ is near the minimum of its
allowed range, and $M_{\rm max}$ is near the maximum of its allowed range, the net
effect of accretion on the masses of all PBHs within the mass spectrum of
our PBH-induced stasis model --- and therefore on the shape of $f_{\rm BH}(M,t)$ --- 
can safely be ignored.  Indeed, we note that all of the results presented in Figs.~5 
and~6 of Ref.~\cite{Dienes:2022zgd} correspond to combinations of model parameters 
for which the negligibility condition $(\Delta M/M)_{\rm max} < 0.01$ is satisfied.

We have shown that the effect of accretion on the masses of the
individual PBHs within our mass spectrum is negligible throughout a significant 
portion of our parameter-space region of interest.  However, while 
$(\Delta M/M)_{\rm max} \ll 1$ is certainly a necessary condition for 
ensuring that the dynamics which give rise to stasis is not disturbed by
accretion, it is not a sufficient one.  We must also verify that the total rate 
$\Gamma_{\rm ac}$ at which energy density is transferred from radiation to matter by 
accretion across the entire cosmological population of PBHs is sufficiently small throughout 
the stasis epoch that it can be neglected.  Indeed, while the rate $\Gamma_e$ at which
energy density is transferred due to evaporation is dominated at an any particular time 
during this epoch by PBHs with masses $M$ only slightly above $\widetilde{{M}}(t)$, 
the fact that $(dM/dt)_{\rm ac}\sim M^2$ implies that PBHs with a far broader range 
of $M$ contribute non-negligibly to $\Gamma_{\rm ac}$.  In general, then, we must require that 
$\Gamma_{\rm ac} \ll \Gamma_e$ throughout this epoch.

In general, the energy-transfer rate due to accretion may be expressed as 
\begin{equation}
  \Gamma_{\rm ac} ~=~ \frac{1}{\rho_{\rm BH}}\int_0^\infty dM\, f_{\rm BH}(M,t) 
    \left(\frac{dM}{dt}\right)_{\rm ac}~.   
\end{equation}
Since our goal is simply to determine the region of our model parameter 
space wherein the condition $\Gamma_{\rm ac} \ll \Gamma_e$ is violated,
we proceed by evaluating $\Gamma_{\rm ac}$ for the simple form of 
$f_{\rm BH}(M,t)$ in Eq.~(\ref{eq:MNote_fMt_fMit_equiv}).  During the 
stasis epoch, $(dM/dt)_{\rm ac}$ is given by Eq.~(\ref{eq:AccretRateGenSimpStasis}),
while 
\begin{equation}
  \frac{a_i^3\rho_{{\rm BH},i}}{a^3\rho_{\rm BH}} ~=~ 
  \frac{\rho_{\rm crit}(t_{\rm PBH})r^3(t_{\rm PBH})}{\rho_{\rm crit}(t)r^3(t)}
    ~=~ \left(\frac{t}{t_{\rm PBH}}\right)^{2\overline{w}/(1+\overline{w})}~.    
\end{equation}
We therefore have
\begin{eqnarray}
  \Gamma_{\rm ac} &~=~& \frac{16\sqrt{3}\overline{w}(\alpha + 1)}
    {(1+\overline{w})^2M_P^2t^2(M_{\rm max}^{\alpha + 1} - M_{\rm min}^{\alpha +1})}
    \nonumber \\ 
    & & ~~~~~~ \times
    \left(\frac{t}{t_{\rm PBH}}\right)^{2\overline{w}/(1+\overline{w})}
    \!\int_{\widetilde{M}(t)}^{M_{\rm max}}\!\!dM_i M_i^{\alpha+1}~.
    \nonumber \\
\end{eqnarray}

By contrast, the energy-transfer rate due to evaporation was shown in
Ref.~\cite{Dienes:2022zgd} to be ${\Gamma_e = -(\alpha+1)/(3t)}$.  Thus, 
through use of Eq.~(\ref{eq:MNote_tPBH}), we obtain an expression 
for the ratio $\Gamma_{\rm ac}/\Gamma_e$ of the form
\begin{eqnarray}
  &&\!\!\!\!\!\!\!\!\frac{\Gamma_{\rm ac}}{\Gamma_e} ~=~
  \frac{144\sqrt{3}\,\overline{w}\epsilon}{(1+\overline{w})^2}
  \left(\frac{t}{t_{\rm PBH}}\right)^{(\overline{w}-1)/(\overline{w}+1)}
  \left(\frac{M_P}{M_{\rm min}}\right)^2
  \nonumber \\ 
  &&\times
  \begin{cases} 
   \displaystyle \frac{1}{\alpha+2}
   \left[\frac{M_{\rm max}^{\alpha+2}-\widetilde{M}^{\alpha+2}(t)}
     {M_{\rm min}(M_{\rm min}^{\alpha+1}-M_{\rm max}^{\alpha+1})}\right] 
     &   -2 < \alpha < -1 \\
  \displaystyle \left(1-\frac{M_{\rm min}}{M_{\rm max}}\right)^{-1}
    \log\left(\frac{M_{\rm max}}{\widetilde{M}(t)}\right)
    &  \alpha = -2~.
  \end{cases}  
  \nonumber \\
\end{eqnarray}

We note that for $0 < \overline{w} < 1/3$ this ratio decreases monotonically 
with $t$. This is primarily due to the fact that the accretion rate is proportional 
to the energy density of the background radiation, which scales like 
$\rho_\gamma \sim t^{-2}$ during stasis, 
whereas $\Gamma_e \sim t^{-1}$ must scale in the same manner with $t$ as does 
$H$ in order for stasis to be maintained.  It will therefore be sufficient for us 
to require that $\Gamma_{\rm ac} \ll \Gamma_e$ at the beginning of stasis, since if
this condition is satisfied when $t = t_{\rm PBH}$ it will be satisfied for all 
$t_{\rm PBH} < t < t_s$ as well.

\begin{figure}[b!]
  \centering
  \includegraphics[width=0.45\textwidth]{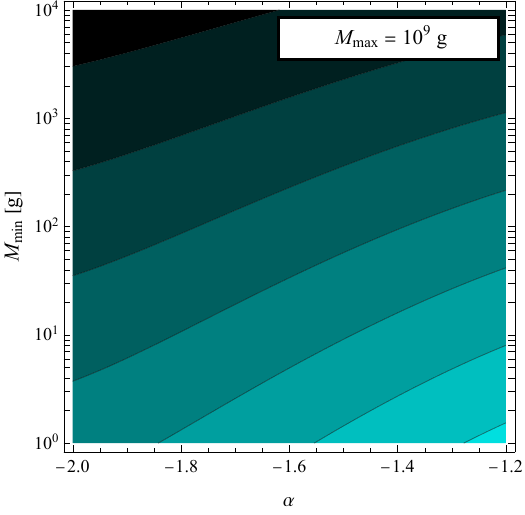}\\
  \includegraphics[width=0.45\textwidth]{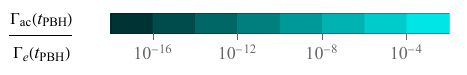}
  \caption{Contours of $\Gamma_{\rm ac}(t_{\rm PBH})/\Gamma_e(t_{\rm PBH})$ 
  within the $(\alpha,M_{\rm min})$-plane at the beginning of the stasis epoch, where
  $\Gamma_{\rm ac}$ denotes the total rate of energy-density transfer from radiation 
  to PBHs from accretion and $\Gamma_e$ denotes the total rate of energy-density transfer 
  from PBHs to radiation from evaporation.  The results shown correspond to the choice of 
  $M_{\rm max} = 10^9$~g --- a choice which yields the maximum value of this ratio 
  subject to the constraint on $M_{\rm max}$ in Eq.~(\ref{eq:MRange}).
  \label{fig:GammaacToGammaeMPlot}}
\end{figure}

In Fig.~\ref{fig:GammaacToGammaeMPlot}, we display contours of
$\Gamma_{\rm ac}(t_{\rm PBH})/\Gamma_e(t_{\rm PBH})$ within the 
$(\alpha,M_{\rm min})$-plane for $M_{\rm max} = 10^9$~g.  (We need not specify 
$\mathcal{N}_{\rm PBH}$, as this ratio is independent of $\mathcal{N}_{\rm PBH}$).  
This choice of $M_{\rm max}$ yields the maximum possible value of 
$\Gamma_{\rm ac}(t_{\rm PBH})/\Gamma_e(t_{\rm PBH})$, given the constraint on 
$M_{\rm max}$ in Eq.~(\ref{eq:MRange}).  We observe that this ratio is much smaller 
than unity throughout the entire region of the $(\alpha,M_{\rm min})$-plane
shown.  Thus, we conclude that the net effect of accretion on the transfer of energy 
density between matter and radiation during stasis is always negligible in comparison 
with the net effect of evaporation. 

In summary, given the results in Figs.~\ref{fig:DeltaMOverMPlot} 
and~\ref{fig:GammaacToGammaeMPlot}, we conclude that the effect of 
accretion both on the masses of the individual PBHs and on the overall 
rate at which energy density is transferred between matter and radiation 
during the stasis epoch is negligible throughout the majority of our 
parameter-space region of interest.  That said, we have also seen that
accretion {\it does}\/ impact the stasis dynamics within certain regions
of that parameter space --- in particular, those in which $\mathcal{N}_{\rm PBH}$
is large or in which the mass spectrum is particularly broad.  Thus, we 
find that accretion yields meaningful constraints on the emergence of 
stasis within the context of PBH-induced stasis scenarios.


\section{Conclusions\label{sec:conclusions}}


In this paper, we have examined how mergers and accretion affect the mass spectrum
of PBHs in scenarios which give rise to cosmological stasis.  While the effect of 
mergers is negligible for PBH masses within the relevant range, we find that 
accretion can have an effect on $f_{\rm BH}(M,t)$.  However, this effect turns out 
to have a significant impact on $f_{\rm BH}(M,t)$ only in cases in which 
$\mathcal{N}_{\rm PBH}$ is large or in which $M_{\rm min}$ and $M_{\rm max}$ lie 
near the opposite endpoints of the phenomenologically allowed range of PBH masses 
specified in Eq.~(\ref{eq:MRange}).  By contrast, throughout the vast majority of the 
parameter-space region of interest for the PBH-induced stasis model we 
have examined here, the effect of accretion is negligible.  This confirms that
this model, the distinctive phenomenological signatures of which were 
discussed in Ref.~\cite{Dienes:2022zgd}, is indeed a viable realization of 
cosmological stasis.  

Notably, these signatures include a number of 
effects on the gravitational-wave spectrum.  Not only does the modification of the 
expansion history alter the profile of the primordial gravitational-wave 
background in a manner such that distortions to the present-day gravitational-wave
background spectrum within different frequency ranges are correlated, but the 
enhancements to the primordial power spectrum which lead to the formation of the PBHs 
themselves also lead to an enhancement in the gravitational-wave background spectrum 
at higher frequencies.  The observation of such a pattern of correlated gravitational-wave
signatures would be compelling evidence that the universe indeed underwent an epoch 
of PBH-induced stasis.  Moreover, any period of matter/radiation stasis leads to an
enhanced growth of density perturbations on small scales~\cite{Dienes:2025tox}, which
can potentially lead to further observational signatures.

Several comments are in order.  First, in this paper we have delineated the regions
of the parameter space of our PBH-induced stasis model within which the stasis 
dynamics is not appreciably impacted by mergers and accretion.  We have not,
however, investigated what happens to $f_{\rm BH}(M,t)$ outside of these regions and what 
the resulting impact is on the expansion history.  An analysis along these lines 
would reveal exactly to what extent the behavior of $\Omega_{\rm BH}$ and $H$ deviates 
from stasis when these processes have a non-negligible impact on $f_{\rm BH}(M,t)$.
We leave such a detailed study for future work.

Second, as discussed in Sect.~\ref{sec:accretion}, there exist significant uncertainties 
regarding the quantitative impact of accretion on the mass of a PBH during the period 
immediately following its production via gravitational collapse during the PBH-formation
epoch.  In this paper, we have assumed that the effect of accretion may either be neglected 
until the quasi-static approximation becomes valid or else may be absorbed via a redefinition 
of our model parameters.  While arguments have been advanced in the literature which 
motivate these assumptions (for a review, see, \eg, Ref.~\cite{Carr:2010wk}), the fact that 
accretion can impact the dynamics which leads to stasis via its impact on the shape of 
$f_{\rm BH}(M,t)$ provides additional impetus for examining the effect that 
accretion has on a PBH during the period immediately following gravitational collapse.

Third, in this paper we have ignored the effect that binary formation has the accretion 
rate --- a phenomenon which was examined in the context of the standard cosmology in
Ref.~\cite{DeLuca:2020qqa}.  Within highly asymmetric binaries, the accretion rate of 
the lighter PBH can be significantly enhanced by the increase in the density of the 
ambient fluid which results from the presence of the heavier PBH.  These effects could 
in principle amplify the distortions which result from accretion, provided that a 
significant fraction of the lighter PBHs are in binaries.  We leave for future work the 
assessment of the quantitative impact that this effect has on the accretion rate for any 
given choice of our model parameters.

Fourth, it has been argued that a quantum effect known as memory 
burden~\cite{Dvali:2018xpy,Dvali:2018ytn} can suppress the evaporation rate of individual 
black holes once some fraction of their initial mass has evaporated~\cite{Dvali:2020wft}. 
The manner in which this effect would impact the cosmological dynamics of our model 
depends on a number of considerations which are subject to significant theoretical 
uncertainties, including the details of how the transition between the semi-classical 
and memory-burdened phases occurs~\cite{Montefalcone:2025akm}.  We therefore leave the 
analysis of this effect for future work.

Fifth, in our analysis, we have focused in our analysis on effects associated with
the homogeneous background cosmology.  While such a treatment is generally sufficient 
for determining whether stasis emerges within a particular scenario and how long it
lasts, given that the stasis phenomenon emerges at the level of the background cosmology, 
we note that in some circumstances additional effects related to spatial inhomogeneities 
in the energy density of a cosmological population of PBHs can modify the manner 
in which $f_{\rm BH}(M,t)$ evolves over time.  One such effect stems from the fact that  
during a protracted era of PBH-domination, 
a perturbation $\delta_{Xk}$ in the energy density of a cosmological matter component $X$ --- 
including the population of PBHs itself --- grows linearly with the scale factor once it enters 
the horizon, whereas it grows only logarithmically 
with $a$ during a radiation-dominated epoch.  This enhanced growth can lead to the efficient 
formation of self-gravitating clusters of PBHs.  Within such structures, PBH binary-capture and
merger rates can be significantly enhanced~\cite{Holst:2024ubt}.  During an epoch 
of matter-radiation stasis, it has likewise been shown~\cite{Dienes:2025tox} that the growth of 
perturbations in a cosmological matter component which is not actively decaying or
evaporating is enhanced.  However, such perturbations do not grow linearly
with $a$ once they enter the horizon, but rather scale with $a$ according to a power law
of the form $\delta_{Xk} \sim a^{q(\overline{w})}$, where the power-law exponent 
$q(\overline{w})$ is determined~\cite{Dienes:2025tox} by the value of $\overline{w}$ and always 
lies within the range $0 < q(\overline{w}) < 1$.  As a result, clusters of PBHs presumably can also 
form during an extended epoch of PBH-induced stasis, although not to the same degree as during a 
PBH-dominated epoch.  Since merger rates are enhanced within these clusters, their 
formation could significantly enhance the impact that mergers have on $f_{\rm BH}(M,t)$ 
during stasis and eventually lead to a destabilization of the stasis dynamics. 
It would be interesting to consider these effects and their impact on the duration of 
stasis in PBH-induced stasis scenarios of this sort.


\section*{Acknowledgements}


The research activities of KRD are supported in part 
by the U.S.\ Department of Energy under Grant DE-FG02-13ER41976 / DE-SC0009913 
and by the U.S.\ National Science Foundation through its employee IR/D program.
The work of LH is funded by the UK Science and Technology Facilities Council 
(STFC) under Grant ST/X000753/1.
The research activities of FH are supported in part by ISF Grant 1784/20, by MINERVA 
Grant 714123, and by MINERVA Grant, Project 7141230301. 
The research activities of TMPT are supported in part by the U.S.\ National Science 
Foundation under Grant PHY-2210283.
The research activities of BT are supported in part by the U.S.\ National Science 
Foundation under Grants PHY-2014104 and PHY-2310622.  
KRD and BT also wish to acknowledge the hospitality of the Center for Theoretical Underground Physics 
and Related Areas (CETUP$^\ast$) and the Institute for Underground Science at 
the Sanford Underground Research Facility (SURF).
The opinions and conclusions expressed herein are those of the 
authors and do not represent any funding agencies. 

\appendix


\section{Probability Densities for Binary Configurations\label{app:BinConfig}}


In this Appendix, we derive the expression for 
$\mathcal{P}_{{\rm n},{\rm nN}}(M_1,M_2,M_3,x,y)$ which appears in 
Eq.~(\ref{eq:MNote_PNnN}).  For pedagogical clarity, we begin by considering the 
idealized case wherein the spatial fluctuations in the primordial energy 
density are described by a Gaussian random field and the spatial distribution of PBHs 
whose masses lie within the range from $M$ to $M+\Delta M$ is therefore Poissonian at 
all scales.  We then examine how the results obtained for this idealized case are modified 
in the more realistic case in which the two-point correlation function $\xi(M,M',x)$ is 
non-vanishing.

For a population of PBHs whose spatial distribution is Poissonian at all scales,
$\Delta M$, the comoving number density of PBHs with masses within the range
$M$ to $M+\Delta M$ is approximately $\Delta n_{\rm PBH} \approx a^3 f_{\rm BH}(M) \Delta M$.  
Thus, the probability that a comoving volume $V$ contains precisely $\nu$ PBHs 
with masses within this range is  
\begin{equation}
  P_\nu(M,\Delta M,V) ~=~ \frac{\big[ V a^3 f_{\rm BH}(M) \Delta M\big]^\nu 
    e^{- V a^3 f_{\rm BH}(M) \Delta M}}{\nu!}~.
  \label{eq:MNote_Poisson}
\end{equation}
Since we shall be particularly interested in comoving volumes which are either spheres 
or spherical shells in what follows, we introduce the compact notation
\begin{equation}
  V(y,x) ~\equiv~ \frac{4\pi}{3}\big(y^3-x^3\big)
  \label{eq:MNote_SphereVol}
\end{equation}
for the volume of a spherical shell of outer radius $y$ and inner radius $x$.  

Our first step in deriving an expression for 
$\mathcal{P}_{{\rm n},{\rm nN}}(M_1,M_2,M_3,x,y)$ in this case is to determine the 
probability that the nearest neighbor of any given PBH of mass $M_1$ is located a comoving 
distance $x$ away from this first PBH and has mass $M_2$, given the probability distribution 
in Eq.~(\ref{eq:MNote_Poisson}).  We begin by dividing the range of possible $M_2$ values into 
bins of width $\Delta M_2$.  We shall label these bins by a discrete index $j = 0,1,2,\ldots$ 
such that $M_{2j}$ denotes the minimum PBH mass in that bin.  The probability that precisely 
one PBH with a mass within the $j$th bin will be located within a thin spherical shell with 
an inner radius $x$ and an outer radius $x + \Delta x$ centered on the location of the first 
PBH is 
\begin{eqnarray}
  && \!\!\!\!\!\! P_1\big(M_{2j},\Delta M_2,V(x+\Delta x,x)\big) ~=~ \nonumber \\ 
  && ~~ \frac{4\pi}{3}a^3 f_{\rm BH}(M_{2j}) \big[(x+\Delta x)^3 - x^3\big]\Delta M_2
    \nonumber \\ 
  && ~~~~ \times \, \exp\left\{-\frac{4\pi}{3}\big[(x+\Delta x)^3 - x^3\big] 
    a^3 f_{\rm BH}(M_{2j}) \Delta M_2\right\}~.\nonumber\\
\end{eqnarray}
By the same token, the probability that no additional PBHs with masses within this same bin 
will be located anywhere within a spherical volume of comoving radius $x$ is
\begin{equation}
  P_0\big(M_{2j},\Delta M_2,V(x,0)\big) ~=~ 
    e^{-4\pi x^3 a^3f_{\rm BH}(M_{2j})\Delta M_2/3}~.
\end{equation}
The joint probability of these two outcomes, which we denote 
$P_{1,0}(j,x,\Delta x)$, is simply the product of these individual probabilities:
\begin{eqnarray}
  P_{1,0}(j,x,\Delta x) &=& 
    \frac{4\pi}{3} a^3f_{\rm BH}(M_{2j}) \big[(x+\Delta x)^3 - x^3\big]\Delta M_2 \nonumber \\ 
    & & ~~ \times\, e^{-4\pi(x+\Delta x)^3 a^3 f_{\rm BH}(M_{2j})\Delta M_2/3}~.~~~~~~~
  \label{eq:MNote_NearestNeighborP_raw}
\end{eqnarray}
For small $\Delta x$, this expression reduces to  
\begin{eqnarray}
  P_{1,0}(j,x,\Delta x) &\,\approx\,& 4\pi  
    e^{-4\pi x^3 a^3f_{\rm BH}(M_2) \Delta M_2/3}\nonumber \\ 
    & & ~~\times \, x^2 a^3f_{\rm BH}(M_2)\Delta M_2 
    \Delta x~.~~~~~
  \label{eq:MNote_NearestNeighborP01}
\end{eqnarray}

The probability $P_{\rm N}(j,x,\Delta x) $ that the nearest neighbor of a given 
PBH will have a mass within the $j$th bin and will be located a comoving distance 
between $x$ and $\Delta x$ away from that other PBH is the infinite product of 
$P_{1,0}(j,x,\Delta x)$ for that particular value of $j$ and the 
$P_0\big(M_{2j'},\Delta M_2, V(x,0)\big)$ for all other $j' \neq j$.  This infinite 
product takes the form
\begin{eqnarray}
  P_{\rm N}(j,x,\Delta x) &\,\approx\,& 4\pi  \left[
     \prod_{j' = 0}^\infty e^{-4\pi x^3 a^3f_{\rm BH}(M_{2j'}) \Delta M_2/3}\right]
     \nonumber \\ 
     & & ~~~~\times \,
    x^2a^3 f_{\rm BH}(M_{2j})\Delta M_2 \Delta x ~.~~~~~~
  \label{eq:MNote_NearestNeighborP_Raw}
\end{eqnarray}
Expressing the product of exponentials as a sum of their arguments and 
taking the limit in which both $\Delta x\to 0$ and $\Delta M_2 \to 0$,
we obtain the differential probability that a given PBH has a nearest neighbor within 
the infinitesimal mass range from $M_2$ to $M_2 + dM_2$ and comoving-distance range 
$x$ to $x+dx$.  Expressing this differential probability in terms of the corresponding 
probability density $\mathcal{P}_{\rm N}(M_2,x)$, we have
\begin{eqnarray}
  && \mathcal{P}_{\rm N}(M_2,x) dM_2 dx ~=~ \nonumber \\ 
  && ~~~~~ 4\pi \exp\left[-\frac{4\pi}{3}x^3a^3\int_0^\infty f_{\rm BH}(M) dM\right]  
    \nonumber \\ 
  && ~~~~~~~~ \times \, x^2 a^3 f_{\rm BH}(M_2) dM_2 dx~.~~~~~~
  \label{eq:MNote_NearestNeighborP}
\end{eqnarray}

Given a particular combination of $M_2$ and $x$ for the nearest neighbor, we also
wish to determine the probability that the next-to-nearest neighbor of a given
PBH has mass $M_3$ and is located a comoving distance $y > x$ away from this first 
PBH.  Proceeding as above, we begin by dividing the range of possible $M_3$ values 
into bins of width $\Delta M_3$ and labeling these bins by a discrete index $j$.
The probability that precisely one PBH with a mass within the $j$th bin 
will be located within a thin spherical shell with inner radius $y$ and outer 
radius $y + \Delta y$ centered on the first PBH is
\begin{eqnarray}
  &&\!\!\!\!\! P_1\big(M_{3j},\Delta M_3, V(y+\Delta y,y)\big) ~=~ \nonumber \\ 
  && ~~\frac{4\pi}{3} a^3f_{\rm BH}(M_{3j}) \Delta M_3 \big[(y+\Delta y)^3 - y^3\big]
    \nonumber \\ 
  && ~~~~ \times \, \exp\left\{-\frac{4\pi}{3}\big[(y+\Delta y)^3 - y^3\big] 
      a^3 f_{\rm BH}(M_{3j}) \Delta M_3\right\}~.\nonumber\\
\end{eqnarray}
The probability that no additional PBHs with masses in this same bin are located 
within a (not necessarily thin) spherical shell with inner radius $x$ and outer 
radius $y$ is 
\begin{equation}
  P_0\big(M_{3j},\Delta M_3, V(y,x)\big) ~=~ 
    e^{-4\pi(y^3 - x^3) a^3f_{\rm BH}(M_{3j}) \Delta M_3/3}~.
\end{equation}
For small $\Delta y$, the joint probability 
$P_{1,0}(j,y,x,\Delta y)$ of these two outcomes is approximately
\begin{eqnarray}
  P_{1,0}\big(j,y,x,\Delta y\big) &\,\approx\,& 
    4\pi e^{-4\pi(y^3 - x^3) a^3f_{\rm BH}(M_{3j}) \Delta M_3/3} 
    \nonumber \\ && ~~\times\, 
    y^2 a^3 f_{\rm BH}(M_{3j}) \Delta M_3 \Delta y~.~~~~~~~~
\end{eqnarray}
The probability $P_{\rm nN}(j,y,x,\Delta y)$  that the next-to-nearest neighbor of 
a given PBH with a nearest neighbor located a comoving distance $x$ away from it
will have a mass within the $j$th bin and will be located a comoving 
distance between $y$ and $y + \Delta y$ away from it is the infinite product of 
$P_{1,0}(j,y,x,\Delta y)$ for that particular value of $j$ and 
the $P_0\big(j',V(x,y)\big)$ for all other $j' \neq j$.  This infinite product
takes the form
\begin{eqnarray}
  P_{\rm nN}(j,y,\Delta y) &\,\approx\,& 4\pi  \left[
     \prod_{j' = 0}^\infty e^{-4\pi (y^3- x^3) a^3 f_{\rm BH}(M_{3j'}) 
       \Delta M_3/3}\right]\nonumber \\ 
     & & ~~~~\times \,
    y^2 a^3 f_{\rm BH}(M_{3j})\Delta M_3 \Delta y ~.~~~~~~
  \label{eq:MNote_NextToNearestNeighborP_Raw}
\end{eqnarray}
Expressing the product of exponentials as a sum of their arguments and 
taking the limit in which both $\Delta y\to 0$ and $\Delta M_3 \to 0$,
we obtain the differential probability that a given PBH has a next-to-nearest neighbor within 
the infinitesimal mass range from $M_3$ to $M_3 + dM_3$ and comoving-distance range 
$y$ to $y+dy$.  Expressing this differential probability in terms of the corresponding 
probability density $\mathcal{P}_{\rm N}(M_3,y)$, we have
\begin{eqnarray}
  && \mathcal{P}_{\rm N}(M_3,y) dM_3 dy ~=~ \nonumber \\ 
  && ~~~~~ 4\pi \exp\left[-\frac{4\pi}{3} (y^3 - x^3) a^3
    \int_0^\infty f_{\rm BH}(M) dM\right] \nonumber \\ 
  && ~~~~~~~~ \times \, y^2 a^3 f_{\rm BH}(M_3) dM_3 dy~.~~~~~~
  \label{eq:MNote_NextToNearestNeighborP}
\end{eqnarray}

The joint probability density $\mathcal{P}_{n,nN}(M_1,M_2,M_3,x,y)$ that the nearest 
neighbor of a particular PBH of mass $M_1$ will be located a comoving distance $x$ away 
from that first PBH, that the next-to-nearest neighbor of that same PBH will be located a 
distance $y$ away from that first PBH, and that these two PBHs will have masses $M_2$ and 
$M_3$, respectively, is the product of the probability densities in 
Eqs.~(\ref{eq:MNote_NearestNeighborP}) and~Eq.~(\ref{eq:MNote_NextToNearestNeighborP}):
\begin{eqnarray}
  && \mathcal{P}_{{\rm n},{\rm nN}}(M_1,M_2,M_3,x,y) ~=~ \nonumber \\ 
  && ~~~~~~ 16\pi^2 x^2 y^2
    \exp\left[-\frac{4\pi}{3}y^3a^3\int_0^\infty f_{\rm BH}(M) dM\right] \nonumber \\ 
  && ~~~~~~~~~~ \times \, a^6 f_{\rm BH}(M_2) f_{\rm BH}(M_3)~.~~~~~~~
  \label{eq:MNote_PNnN_Raw}
\end{eqnarray}
For a population of PBHs which are perfectly Poisson-distributed, the two-point correlation
function is simply $\xi(M,M',x) = 0$, as discussed above.  Thus, for such a population of 
PBHs, the expression in Eq.~(\ref{eq:MNote_Nofy}) reduces to
\begin{equation}
  N(M',y) ~=~ \frac{4\pi}{3}y^3a^3 \int_0^\infty f_{\rm BH}(M) dM~,  
  \label{eq:MNote_Nofy_Poisson}
\end{equation}
which is independent of the value of $M'$.  Thus, for such a population of PBHs, we may 
express $\mathcal{P}_{{\rm n},{\rm nN}}(M_1,M_2,M_3,x,y)$ in terms of the quantities 
defined in Eqs.~(\ref{eq:MNote_dNdMdx}) and~(\ref{eq:MNote_Nofy}) as
\begin{eqnarray}
  && \!\!\!\!\! \mathcal{P}_{{\rm n},{\rm nN}}(M_1,M_2,M_3,x,y) ~=~ \nonumber \\
  && ~~~     e^{-N(M_1,y)} \frac{\partial^2 N(M_2,M_1,x)}{\partial M_2 \partial x}      
       \frac{\partial^2 N(M_3,M_1,y)}{\partial M_3 \partial y}~,~~~~~~~~
  \label{eq:MNote_PNnN_app}
\end{eqnarray}
which is independent of $M_1$. 

While the expression in Eq.~(\ref{eq:MNote_PNnN_app}) is strictly valid only for a perfectly 
spatially homogeneous, Poisson-distributed population of PBHs, this result can be generalized 
in a straightforward manner to the case in which the correlation function $\xi(M,M',x)$ is 
non-vanishing~\cite{Ballesteros:2018swv}.  The corresponding expression for the 
joint probability density in this case may be obtained by replacing the expression 
for $V(y,x)$ in Eq.~(\ref{eq:MNote_SphereVol}) with
\begin{equation}
  V(M,M',y,x) ~\equiv~ 4\pi\int_x^y \big[1+\xi(M,M',z)\big] z^2 dz~.
\end{equation}
One can then derive an expression for $\mathcal{P}_{{\rm n},{\rm nN}}(M_1,M_2,M_3,x,y)$ in
a manner analogous to the manner in which we have derived the corresponding result for the 
purely Poisson case.  Noting that the fundamental theorem of calculus implies that
\begin{equation}
  \lim_{\Delta x\to 0} V(M,M',x+\Delta x,x) ~=~ 
    4\pi \big[1+\xi(M,M',x)\big] x^2 dx
\end{equation}
for a spherical shell of infinitesimal width, we find that 
$\mathcal{P}_{{\rm n},{\rm nN}}(M_1,M_2,M_3,x,y)$ is given by an expression of exactly 
the same form as the expression Eq.~(\ref{eq:MNote_PNnN_app}), but in which
$\partial N(M_2,M_1,x)/dM_2dx$ and $\partial N(M_3,M_1,x)/dM_2dx$
include the full expressions for $\xi(M_2,M_1,x)$ and $\xi(M_3,M_1,y)$,
respectively, and in which the expression in Eq.~(\ref{eq:MNote_Nofy_Poisson}) is 
modified to
\begin{eqnarray}
  N(M',y) &~=~& 4\pi a^3 \int_0^y dx\, x^2 \int_0^\infty dM \Big\{ f_{\rm BH}(M) 
    \nonumber \\
    & &~~~~~~~~\times \,  \big[1+\xi(M,M',x)\big]\Big\}~,~~~~   
\end{eqnarray}
which has the form as the expression in Eq.~(\ref{eq:MNote_Nofy}) for non-vanishing
$\xi(M,M',x)$.

\bibliography{references}
\end{document}